\documentclass[preprint,12pt]{elsarticle}

\usepackage{amssymb}

\usepackage{amsmath}
\usepackage{graphicx}
\usepackage{txfonts}

\usepackage{multirow}
\usepackage{multicol}
\usepackage{makecell}
\usepackage{subfigure}

\usepackage{flushend,cuted}

\journal{Arvix}

\begin{document}

\begin{frontmatter}



\title{Bifurcation mechanism of quasihalo orbit from Lissajous Orbit}


\author[inst1]{Mingpei Lin\corref{mycorrespondingauthor}}
\cortext[mycorrespondingauthor]{: lin.mingpei.d2@tohoku.ac.jp}

\affiliation[inst1]{organization={Advanced Institute for Materials Research, Tohoku University},
            addressline={2-1-1 Katahira, Aoba-ku}, 
            city={Sendai},
            postcode={980-8577}, 
            country={Japan}}

\author[inst1]{Hayato Chiba}

\begin{abstract}
This paper presents a general analytical method to describe the center manifolds of collinear libration points in the Restricted Three-body Problem (RTBP). It is well-known that these center manifolds include Lissajous orbits, halo orbits, and quasihalo orbits. Previous studies have traditionally tackled these orbits separately by iteratively constructing high-order series solutions using the Lindstedt-Poincaré method.
Instead of relying on resonance between their frequencies, this study identifies that halo and quasihalo orbits arise due to intricate coupling interactions between in-plane and out-of-plane motions.
To characterize this coupling effect, a novel concept, coupling coefficient $\eta$, is introduced in the RTBP, incorporating the coupling term $\eta \Delta x$ in the $z$-direction dynamics equation, where $\Delta$ represents a formal power series concerning the amplitudes. Subsequently, a uniform series solution for these orbits is constructed up to a specified order using the Lindstedt-Poincaré method.
For any given paired in-plane and out-of-plane amplitudes, the coupling coefficient $\eta$ is determined by the bifurcation equation $\Delta = 0$. When $\eta$ = 0, the proposed solution describes Lissajous orbits around libration points. As $\eta$ transitions from zero to non-zero values, the solution describes quasihalo orbits, which bifurcate from Lissajous orbits. Particularly, halo orbits bifurcate from planar Lyapunov orbits if the out-of-plane amplitude is zero. The proposed method provides a unified framework for understanding these intricate orbital behaviors in the RTBP.
\end{abstract}



\begin{keyword}
Restricted three-body problem \sep Center manifold \sep Lissajous orbit 
\sep Quasihalo orbit \sep Coupling coefficient 

\end{keyword}

\end{frontmatter}


\section{Introduction}
\label{sec:introduction}
The Restricted Three-body Problem (RTBP) serves as a fundamental dynamical model to study the motion of asteroids or satellites under the gravitational influence of two primaries \citep{conley1968low,howell1997application,lara2022perturbation}. The RTBP has five equilibrium points called the Lagrange points $ L_i$ ($i$ = 1, 2, …, 5), among which three so-called collinear libration points are unstable, and the remaining two are triangular libration points. The linear behavior of three collinear libration points is of the type center × center × saddle. The center manifold of three collinear libration points includes Lissajous orbits around libration points, halo orbits, and quasihalo orbits around halo orbits \citep{gomez2001dynamics}. These libration point orbits play a key role in astronomy and many space science missions \citep{richardson1980halo,addington1995dual,bosanac2019trajectory,duan2019orbit}. 

Numerical methods (numerical integration, optimization technology) and analytical methods (parameterization method, norm form, Lindstedt-Poincaré method) are often used to study the dynamics in the center manifolds of collinear libration points. In most cases, numerical methods still require accurate analytical initial values. On the other hand, high-order analytical solutions explicitly and accurately describe the dynamics around libration points, which is very useful for the preliminary space mission design. Farquhar initially proposed the third-order solution for small-amplitude Lissajous orbits and introduced the concept of “halo orbit” \citep{farquhar1969control,farquhar1973quasi}. Richardson \cite{richardson1980analytic} presented a third-order analytical solution of halo orbits around collinear libration points in the RTBP. He employed the Lindstedt-Poincaré method to construct this solution by introducing a correction term to modify the out-of-plane frequency. By means of the Lindstedt-Poincaré method, Jorba and Masdemont \cite{jorba1999dynamics,jorba1999nonlinear} obtained high-order series solutions of Lissajous orbits and halo orbits around collinear libration points with the help of ad hoc algebraic manipulators. Celletti and Luo et al. \cite{celletti2015lissajous, luo2020lissajous} conducted an analytical study of the Lissajous and halo orbits around collinear points in RTBP for arbitrary mass ratios, utilizing a canonical transformation procedure. Paez and Guzzo \cite{paez2022semi} provided a semi-analytical construction of halo orbits and halo tubes in the elliptic RTBP by implementing a nonlinear Floquet–Birkhoff resonant normal form. However, these analytical methods cannot yield high-order series solutions of quasihalo orbits. Gómez et al. \cite{gomez1998quasihalo} took a halo orbit with a constant amplitude as a backbone (periodic-varying equilibrium point), and further employed Floquet transformation and Lindstedt-Poincaré method to construct quasihalo orbits. This method necessitates specifying the magnitude of the reference halo orbit and using Floquet transformation, and does not directly provide a complete analytical expression for quasihalo orbits.".

So far, there is no semi-analytical method to uniformly describe Lissajous orbits, halo orbits and quasihalo orbits in the center manifolds, although some numerical methods, such as norm form \cite{jorba1999dynamics} and multiple Poincaré section \citep{kolemen2012multiple}, can provide a comprehensive description for the center manifolds of RTBP. References \citep{richardson1980analytic,jorba1999dynamics,jorba1999nonlinear} successfully gave the third or higher-order series solutions of the Lissajous and halo orbits, yet they failed to construct a high-order series solution of the quasihalo orbits. Since the discovery of halo orbits in 1969, the astrodynamics community has traditionally attributed their origin to bifurcations from planar Lyapunov orbits when the in-plane and out-of-plane frequencies are in a 1:1 resonance. In this paper, we also contend that halo orbits indeed arise from bifurcations of planar Lyapunov orbits. However, We employ a new mechanism, coupling-induced bifurcation, to explain the generation of halo and quasihalo orbits. Consider a two-degree-of-freedom nonlinear dynamical system similar to the part of the center manifolds in the RTBP, but in this system, the degrees of freedom are decoupled, i.e., ${\left[ {\ddot x,\ddot y} \right]^{\rm{T}}} = {\left[ {f(x),f(y)} \right]^{\rm{T}}}$. It is evident that bifurcations of periodic orbits do not occur even under frequency resonance. Drawing a distinction between this hypothetical dynamical system and the RTBP, it can be inferred that the true trigger for the emergence of halo orbits through planar Lyapunov orbit bifurcation is the nonlinear coupling effect between the two degrees of freedom, rather than a resonance between the two frequencies. Considering the nonlinear coupling of in-plane and out-of-plane motions, this paper introduce the concept of a coupling coefficient $\eta$ into the RTBP for the first time. Additionally, a new correction term $\eta \Delta x$ is incorporated into the RTBP equation to construct a unified semi-analytical solution for the center manifolds of collinear libration points in the RTBP. Here, $\Delta$ is a power series concerning in-plane and out-of-plane amplitudes. When the coupling coefficient $\eta$ is zero, the correction item is inactive, and the series solution describes Lissajous orbits around libration points. Otherwise, the series solution describes quasihalo orbits, of course, including halo orbits. In this way, the center manifolds of collinear libration points in the RTBP are uniformly described with a high-order series solution.

The contribution of this paper lies in the novel proposition of the coupling-induced bifurcation mechanism to explain the generation of quasihalo orbits. This marks the first complete approximate analytical solution for quasihalo orbits, providing a comprehensive analytical solution for the entire central manifold, including Lissajous orbits, halo orbits, and quasihalo orbits. Additionally, the practical convergence domain of quasihalo orbits is presented for the first time. Lastly, the proposed method transforms the dynamic orbit bifurcation problem of vector fields into a static bifurcation problem for the solutions of the bifurcation equation. This method is applicable for general analytical bifurcation analysis in dynamical systems.

The remainder of this paper is organized as follows. Section 2 introduces the dynamical model of the RTBP. In section 3, a general analytical construction method for the center manifolds of collinear libration points in the RTBP is presented. Results and discussions are provided in Section 4. Finally, Section 5 makes some concluding remarks.

\section{Dynamical model}
This section introduces the dynamical model of the classical RTBP. This model serves as a good approximation for the motion of an infinitesimal particle (asteroids or spacecraft) under the gravitational attraction of two primaries. The attraction of the particle on the primaries is neglected so the two primaries rotating around their common center of mass in a Kepler orbit. Here, our focus lies on the circular orbit, i.e., a Kepler orbit with zero eccentricity. 
The motion of the particle is usually described in a synodic coordinate system. In this system, the origin is positioned at the centroid of the two primaries. The $X$-axis points from the smaller primary to larger primary, the $Z$-axis is perpendicular to the plane of the Kepler orbit and positive in the direction of the angular momentum, and the $Y$-axis completes a right-hand triad. Consequently, the position vector of the smaller primary and larger primary is ($-\mu$, 0, 0) and ($1-\mu$, 0, 0), respectively. Then, the differential equation governing the motion of the particle in the synodic coordinate system is expressed as follows \cite{koon2000dynamical}: 
\begin{equation}
\begin{aligned}
\ddot X - 2\dot Y &= \frac{{\partial \Omega }}{{\partial X}}\\
\ddot Y + 2\dot X &= \frac{{\partial \Omega }}{{\partial Y}}\\
\ddot Z &= \frac{{\partial \Omega }}{{\partial Z}}
\end{aligned}
\label{eq:dynamical model}
\end{equation}
with
\begin{equation}
\Omega \left( {X,Y,Z} \right) = \frac{1}{2}\left( {{X^2} + {Y^2}} \right) + \frac{{1 - \mu }}{{{r_1}}} + \frac{\mu }{{{r_2}}} + \frac{1}{2}\mu \left( {1 - \mu } \right)
\label{eq:OMG}
\end{equation}
where $\mu = m_2 / (m_1 + m_2)$ is the mass parameter of the system. $m_1$ and $m_2$ are the mass of the smaller primary and larger primary, respectively. $r_1$ and $r_2$ are the distance from the particle to the smaller primary and larger primary, respectively. 
\begin{equation}
\begin{array}{l}
{r_1}^2 = {\left( {X + \mu } \right)^2} + {Y^2} + {Z^2}\\
{r_2}^2 = {\left( {X - 1 + \mu } \right)^2} + {Y^2} + {Z^2}
\end{array}
\label{eq:r1r2}
\end{equation}
This model has a Jacobi integral 
\begin{equation}
C = 2\Omega  - ({\dot X^2} + {\dot Y^2} + {\dot Z^2})
\label{eq:Jacobi}
\end{equation}
As mentioned in the Introduction section, the RTBP has five equilibrium points. In this paper, we focus on the three collinear libration points $L_1$, $L_2$, and $L_3$. Let $\gamma_i (i =1, 2,3)$ denote the distance from Li to the closet primary. This distance is determined by the unique positive root of the Euler quantic equation \citep{gomez2001dynamics,koon2000dynamical},
\begin{equation}
\begin{array}{l}
\gamma _i^5 \mp (3 - \mu )\gamma _i^4 + (3 - 2\mu )\gamma _i^3 - \mu \gamma _i^2 \pm 2\mu {\gamma _i} - \mu  = 0,{\rm{  }}\qquad \qquad \qquad \quad i = 1,2\\
\gamma _i^5 + (2 + \mu )\gamma _i^4 + (1 + 2\mu )\gamma _i^3 - (1 - \mu )\gamma _i^2 - 2(1 - \mu ){\gamma _i} - (1 - \mu ) = 0, \ i = 3
\end{array}
\label{eq:gamm}
\end{equation}
When focusing on the selected libration point $L_i$, a coordinate transformation is performed to move the origin of coordinate system to the libration point $L_i$,
\begin{equation}
\begin{array}{l}
X =  - {\gamma _i}x + \mu  - 1 + {\gamma _i},Y =  - {\gamma _i}y,Z = {\gamma _i}z,{\rm{  }} \  i = 1,2\\
X = {\gamma _i}x + \mu  + {\gamma _i},Y = {\gamma _i}y,Z = {\gamma _i}z,\quad \quad \quad {\rm{  }}i = 3
\end{array}
\label{eq:X_transform}
\end{equation}
Then, the dynamical model (\ref{eq:dynamical model}) of the RTBP can be reformulated as 
\begin{equation}
\begin{aligned}
\ddot x - 2\dot y &= \frac{{\partial \Omega }}{{\partial x}}\\
\ddot y + 2\dot x &= \frac{{\partial \Omega }}{{\partial y}}\\
\ddot z &= \frac{{\partial \Omega }}{{\partial z}}
\end{aligned}
\label{eq:dynamical model2}
\end{equation}
where
\begin{equation}
\begin{aligned}
\Omega \left( {x,y,z} \right) &= \frac{1}{2}\left( {{{\left( {\mu  - 1 \mp \gamma \left( {x - 1} \right)} \right)}^2} + {\gamma ^2}{y^2}} \right) \\
    &+ \frac{{1 - \mu }}{{{r_1}}} + \frac{\mu }{{{r_1}}} + \frac{1}{2}\mu \left( {1 - \mu } \right)
\end{aligned}
	\label{eq:OMG2}
\end{equation}

In order to construct a high-order series solution for the center manifolds in the RTBP in the following section, the motion equation (\ref{eq:dynamical model2}) is expanded in power series using the Legendre polynomials \citep{jorba1999dynamics,richardson1980note}, 
\begin{equation}
\begin{aligned}
\ddot x - 2\dot y - (1 + 2{c_2})x &= \frac{\partial }{{\partial x}}\sum\limits_{n \ge 3} {{c_n}{\rho ^n}{P_n}\left( {\frac{x}{\rho }} \right)} \\
\ddot y + 2\dot x + ({c_2} - 1)y &= \frac{\partial }{{\partial y}}\sum\limits_{n \ge 3} {{c_n}{\rho ^n}{P_n}\left( {\frac{y}{\rho }} \right)} \\
\ddot z + {c_2}z &= \frac{\partial }{{\partial z}}\sum\limits_{n \ge 3} {{c_n}{\rho ^n}{P_n}\left( {\frac{z}{\rho }} \right)} 
\end{aligned}
	\label{eq:dynamical model_Legendre}
\end{equation}
where $\rho = x^2 + y^2 + z^2$, $P_n$ is Legendre polynomials, and the constant coefficients $c_n(\mu)$ are depend on the system parameters $\mu$,
\begin{equation}
\begin{array}{l}
{c_n}(\mu ) = \frac{1}{{\gamma _i^3}}\left( {{{( \pm 1)}^n} + {{( - 1)}^n}\frac{{(1 - \mu )\gamma _i^{n + 1}}}{{{{(1 \mp \gamma )}^{n + 1}}}}} \right),\ {\rm{  for }} \ {L_i},\ i = 1,2\\
{c_n}(\mu ) = \frac{{{{( - 1)}^n}}}{{\gamma _i^3}}\left( {1 - \mu  + \frac{{\mu \gamma _i^{n + 1}}}{{{{(1 + \gamma )}^{n + 1}}}}} \right),\quad\quad {\rm{ for }} \ {L_i},\ i = 3.
\end{array}
	\label{eq:cn}
\end{equation}
The left-hand linear terms form the foundational components of the series solution, while the right-hand nonlinear ones are crucial for constructing halo and quasihalo orbits. They are well-known as bifurcations from large-amplitude planner Lyapunov orbit.

\section{Analytical construction of center manifolds in RTBP}\label{sect:construction}
In this section, a semi-analytical solution for the center manifolds of collinear libration points in the RTBP is constructed using the Lindstedt–Poincaré method. Lindstedt–Poincaré method is an iterative computational technology that begins from the fundamental first-order solution of the system. Subsequently, it continuously adjusts the relationship between the frequency and amplitude to obtain a higher-order series solution by iterating the known low-order solution step by step. 
\subsection{Form of the analytical solution}
First, the first-order solution for the center manifolds of collinear libration points in the RTBP should be found. It is well known that the center manifolds include Lissajous orbits, halo orbits, and quasihalo orbits around halo orbits. For Lissajous orbits, the first-order solution naturally arises by solving the linear part of (\ref{eq:dynamical model_Legendre}),
\begin{equation}
    \begin{aligned}
    \ddot x - 2\dot y - (1 + 2{c_2})x &= 0\\
    \ddot y + 2\dot x + ({c_2} - 1)y &= 0\\
    \ddot z + {c_2}z &= 0.
    \end{aligned}
	\label{eq:linear equation}
\end{equation}
The solution of (\ref{eq:linear equation}) is 
\begin{equation}
\begin{array}{l}
x(t) = {\rm{ }}\alpha \cos ({\omega _0}t + {\varphi _1})\\
y(t) = \kappa \alpha \sin ({\omega _0}t + {\varphi _1})\\
z(t) = \beta \cos ({v_0}t + {\varphi _2})
\end{array}
	\label{eq:linear solution}
\end{equation}
where 
\begin{equation}
{\omega _0} = \sqrt {\frac{{2 - {c_2} + \sqrt {9c_2^2 - 8{c_2}} }}{2}},
{v_0} = \sqrt {{c_2}},
\kappa  =  - \frac{{{\omega _0}^2 + 1 + {c_2}}}{{2{\omega _0}}}
	\label{eq:frequencies}
\end{equation}
$\alpha$ and $\beta$ represent the in-plane and out-of-plane amplitudes, respectively. $\phi _1$ and $\phi _2$ denote the corresponding phases. 
As stated in the Introduction section, halo orbits bifurcate from the planar Lyapunov periodic orbits without requiring the in-plane and out-of-plane frequencies to be equal. This phenomenon is primarily due to the nonlinear coupling of in-plane and out-of-plane motions in RTBP. Considering the coupling between in-plane and out-of-plane motions, a natural concept of coupling coefficient $\eta$ can be defined to characterize the degree of coupling between the linear motion in plane and out of plane. Consequently, the basic first-order solution can be reformulated as
\begin{equation}
\begin{array}{l}
x(t) = {\rm{ }}\alpha \cos ({\omega _0}t + {\varphi _1})\\
y(t) = \kappa \alpha \sin ({\omega _0}t + {\varphi _1})\\
z(t) = \eta \alpha \cos ({\omega _0}t + {\varphi _1}) + \beta \cos ({v_0}t + {\varphi _2})
\end{array}
	\label{eq:modified linear solution}
\end{equation}
To derive the solution (\ref{eq:modified linear solution}) for (\ref{eq:linear equation}), it is necessary to introduce a correction term and rewrite it as
\begin{equation}
\begin{aligned}
\ddot x - 2\dot y - (1 + 2{c_2})x &= 0\\
\ddot y + 2\dot x + ({c_2} - 1)y &= 0\\
\ddot z + {c_2}z &= \eta {d_{00}}x
\end{aligned}
	\label{eq:modified linear equation}
\end{equation}
where $d_{00} = c_2 - {\omega_0}^2 \ne 0$ is a constant correction factor and $\eta d_{00} = 0$. It is evident that for linear equation (\ref{eq:modified linear equation}), the only trivial solution is $\eta = 0$, indicating the absence of coupling between the in-plane and out-of-plane motions. Halo/quasihalo orbits only appears when the coupling coefficient $\eta$ is not-zero in higher-order solutions. Therefore, we extend (\ref{eq:modified linear equation}) to higher-order case by introducing a higher-order correction term to the third equation of (\ref{eq:dynamical model_Legendre}) as follows:
\begin{equation}
\begin{aligned}
\ddot x - 2\dot y - (1 + 2{c_2})x &= \frac{\partial }{{\partial x}}\sum\limits_{n \ge 3} {{c_n}{\rho ^n}{P_n}\left( {\frac{x}{\rho }} \right)} \\
\ddot y + 2\dot x + ({c_2} - 1)y &= \frac{\partial }{{\partial y}}\sum\limits_{n \ge 3} {{c_n}{\rho ^n}{P_n}\left( {\frac{y}{\rho }} \right)} \\
\ddot z + {c_2}z &= \frac{\partial }{{\partial z}}\sum\limits_{n \ge 3} {{c_n}{\rho ^n}{P_n}\left( {\frac{z}{\rho }} \right)}  + \eta \Delta x
\end{aligned}
	\label{eq:modified dynamical equation}
\end{equation}
where the higher-order term $\eta \Delta x$ is the product of the coupling coefficient $\eta$, the nonlinear correction factor $\Delta$, and $x$, satisfying $\Delta  = \sum\limits_{} {{d_{ij}}{\alpha ^i}{\beta ^j}} $ and $\eta \Delta = 0$. During the process, non-trivial solutions of $\eta$ are determined from the condition $\Delta = 0$. When the nonlinear terms in (\ref{eq:modified dynamical equation}) are taken into account, the semi-analytical solution is formulated as formal expansions in powers of $\alpha$ and $\beta$,
\begin{equation}
\begin{array}{l}
x\left( t \right) = \sum\limits_{i + j \ge 1} {\left( {\sum\limits_{\left| k \right| \le i,\left| m \right| \le j} {{x_{ijkm}}\cos \left( {k{\theta _1} + m{\theta _2}} \right)} } \right){\alpha ^i}{\beta ^j}} \\
y\left( t \right) = \sum\limits_{i + j \ge 1} {\left( {\sum\limits_{\left| k \right| \le i,\left| m \right| \le j} {{y_{ijkm}}\sin \left( {k{\theta _1} + m{\theta _2}} \right)} } \right){\alpha ^i}{\beta ^j}} \\
z\left( t \right) = \sum\limits_{i + j \ge 1} {\left( {\sum\limits_{\left| k \right| \le i,\left| m \right| \le j} {{z_{ijkm}}\cos \left( {k{\theta _1} + m{\theta _2}} \right)} } \right){\alpha ^i}{\beta ^j}} 
\end{array}
	\label{eq:high-order solution}
\end{equation}
where ${\theta _1}{\rm{ = }}{\omega _{\rm{ }}}t + {\varphi _1}$,${\theta _2}{\rm{ = }}\nu t + {\varphi _2}$. Considering the nonlinear terms, the frequencies should also be expanded in the power series of $\alpha$ and $\beta$,
\begin{equation}
\omega  = \sum\limits_{i,j \ge 0} {{\omega _{ij}}} {\alpha ^i}{\beta ^j},v = \sum\limits_{i,j \ge 0} {{v_{ij}}} {\alpha ^i}{\beta ^j}
	\label{eq:high-order frequencies}
\end{equation}
Moreover, we have the constraint 
\begin{equation}
\eta \Delta  = \eta \sum\limits_{i + j = n} {{d_{ij}}{\alpha ^i}{\beta ^j}}  = 0
	\label{eq:high-order coupling}
\end{equation}
which provides the implicit relationship between $\eta$ and $\alpha$ and $\beta$, i.e., $\eta = \eta (\alpha,\beta)$. Equations (\ref{eq:high-order solution}), (\ref{eq:high-order frequencies}), and (\ref{eq:high-order coupling}) together constitute a comprehensive analytical description of the central manifolds around the collinear libration in the RTBP.

\textit{Remark 1}. In fact, the above correction is not unique, as long as the correction term can represent the coupling between in-plane and out-of-plane motions, such as adding $\eta \Delta y$ to the third equation or $\eta \Delta z$ to the first (second) equation of (\ref{eq:modified dynamical equation}). 

\textit{Remark 2}. When $\eta = 0$, the solution (\ref{eq:high-order solution}) represents Lissajous orbits. Specifically, if $\beta = 0$, it corresponds to planar Lyapunov orbits; if $\alpha = 0$, it corresponds to vertical Lyapunov orbits. When $\eta \ne 0$, the solution (\ref{eq:high-order solution}) represents quasihalo orbit. Particularly, if $\eta > 0$ $(\eta < 0)$ and $\beta = 0$, it results in north (south) halo orbits. No solutions exist if $\alpha = 0$, which will be demonstrated in the next section. In sum up, the high-order series solution (\ref{eq:high-order solution}) uniformly describes the center manifolds of collinear libration points in the RTBP.

\textit{Remark 3}. In (\ref{eq:high-order solution}) and (\ref{eq:high-order frequencies}), $i$ and $j \in \mathbb{N}$, $k$ and $m \in \mathbb{Z}$. Due to the symmetry of the RTBP, $x(t)$ and $z(t)$ are formulated as a cosine series, and $y(t)$ as a sine series. $p$ and $q$ have the same parity as $i$ and $j$, respectively. Besides, due to the symmetry of sine and cosine functions, it can be assumed that $m \ge 0$, and $n \ge 0$ when $m = 0$. The series of $\omega$ and $v$ only include even items. These properties are useful for conserving computational storage and time.

\subsection{Solving for undetermined coefficients}
Now, our goal is to compute the coefficients $x_{ijkm}$, $y_{ijkm}$, $z_{ijkm}$, $\omega_{ij}$, $v_{ij}$ in (\ref{eq:high-order solution}) and  (\ref{eq:high-order frequencies}) up to a finite order $n$. The Lindstedt-Poincaré method is utilized to calculate these coefficients following an iterative scheme from the linear solution. Compared to the solution of the linear part (\ref{eq:modified linear solution}), we can determine $x_{1010} = 1$, $y_{1010} = \kappa$, $z_{1010} = \eta$, $z_{0101} = 1$, $\omega_{00} = \omega_0$, $v_{00} = v_0$. By substituting this linear solution into (\ref{eq:modified dynamical equation}), the coefficients of the second-order solution can be derived. Similarly, when the coefficients up to order $n - 1$ are obtained, i.e., $x(t)$, $y(t)$, and $z(t)$ are determined up to order $n - 1$, $\omega$ and $v$ are determined up to order $n - 2$. Substituting them into the right side of (\ref{eq:modified dynamical equation}), we can obtain three power series up to order $n$, denoted by $p$, $q$, and $r$. Here, what we are interested in are those $n$-order terms. Without losing generality, the $n$-order terms of $p$, $q$, and $r$ are denoted by $p_{ijkm}$, $q_{ijkm}$, and $r_{ijkm}$ ($i + j = n$) respectively.

\begin{table}[h]
	\centering
    \tiny
	\caption{Derivatives of $x$ and $y$ with respect to time and $\Delta x$.}
	\label{tab:derivatives}
	\begin{tabular}{cccccccc} 
		\hline
		& $f$ & $g$ & $\omega(\partial x / \partial {\theta_1})$ &
          $v(\partial x / \partial {\theta_2})$ & $\omega(\partial y / \partial {\theta_1})$ & $v(\partial y / \partial {\theta_2})$ & $\Delta x$\\
		\hline
		\multirow{2}{*}{\makecell{ Unknown \\ term} } & 0 & n & $-\omega_{00} kx_{ijkm}$ & 
        $-v_{00} mx_{ijkm}$ & $\omega_{00} ky_{ijkm}$ &  $v_{00} my_{ijkm}$ &
        $d_{00} x_{ijkm}$ \\
		   & $n-1$ & 1 & $-\omega_{i-1j} k\delta_{1k0m}$ & 0 & $\kappa \omega_{i-1j}k\delta_{1k0m}$ & 0 & $d_{i-1j}\delta_{1k0m}$ \\
        \hline
		Known term & \makecell{1,2,...,\\ $n-2$} & $n-(ij)_f$ & $-\omega_{(ij)_f} kx_{(ij)_gkm}$ & $-v{(ij)_f} mx_{(ij)_gkm}$ & $\omega_{(ij)_f} ky_{(ij)_gkm}$ & $v{(ij)_f} my_{(ij)_gkm}$ & ${d_{{{(ij)}_f}}}{x_{{{(ij)}_g}km}}$\\
		\hline
	\end{tabular}
\end{table}

\begin{table*}[h]
	\centering
     \tiny
	\caption{Second derivatives of $x$, $y$, and $z$ with respect to time.}
	\label{tab:second derivatives}
	\begin{tabular}{cccccc} 
		\hline
  & $f$ & $g$ & $\omega^2 (\partial^2[x,y,z]/\partial \theta_1^2)$ 
  & $2\omega v (\partial^2[x,y,z]/\partial \theta_1 \theta_2)$
  & $v^2 (\partial^2[x,y,z]/\partial \theta_1^2)$ \\
        \hline
        \multirow{2}{*}{Unknown term} & 0 & $n$ & $-\omega_{00}^2k^2[x,y,z]_{ijkm}$ & $-2\omega_{00}v_{00}km[x,y,z]_{ijkm}$ & $- v_{00}^2 m^2 [x,y,z]_{ijkm}$\\ 
        & $n-1$ & 1 & $\begin{array}{c} -\Omega_{i-1j}\delta_{1k}\delta_{0m} \\
        -\Omega_{i-1j}\kappa\delta_{1k}\delta_{0m} \\ -\Omega_{i-1j}\eta\delta_{1k}\delta_{0m}\end{array}$ & 0 & $\begin{array}{c} 0 \\ 0 \\ -N_{(ij-1)}\delta_{0k}\delta_{1m} \end{array}$ \\
        \hline
        Known term & \makecell{1,2,..., \\ $n-2$} & $n - (ij)_f$ & $-\Omega_{(ij)_f}k^2[x,y,z]_{(ij)_gkm}$ & $-2\Omega_{(ij)_{f_1}}N_{(ij)_{f_2}}km[x,y,z]_{(ij)_gkm}$ & $-N{(ij)_f}m^2[x,y,z]_{(ij)_gkm}$\\
        \hline
	\end{tabular}
\end{table*}

Next, we analyze the composition of $n$-order terms on the left side of (\ref{eq:modified dynamical equation}). According to (\ref{eq:high-order solution}), the derivatives of variable $x$ can be expressed as
\begin{equation}
\begin{array}{l}
\dot x = \omega \frac{{\partial x}}{{\partial {\theta _1}}} + \nu \frac{{\partial x}}{{\partial {\theta _2}}}\\
\ddot x = {\omega ^2}\frac{{{\partial ^2}x}}{{\partial \theta _1^2}} + 2\omega v\frac{{{\partial ^2}x}}{{\partial {\theta _1}\partial {\theta _2}}} + {v^2}\frac{{{\partial ^2}x}}{{\partial \theta _2^2}}
\end{array}
	\label{eq:first derivative}
\end{equation}
Similarly, we can obtain the derivatives of $y$ and $z$. Let $fg$ denote the first-order derivative term, where $f$ represents the power series of the frequencies ($\omega$ and $v$), and $g$ represents the coordinate variables ($x$, $y$, and $z$). $(ij)_f$ and $(ij)_g$ denote their corresponding order. Then, the $fg$ satisfying $(ij)_f + (ij)_g = n$ constitutes the $n$-order terms we require. When $(ij)_f$ is 0 or $n - 1$, the corresponding $(ij)_g$ is $n$ or 1 and $f g$ is an unknown term that needs to be solved. 
When $(ij)_f = 1, 2, ..., n - 2$, $f g$ is a known term that needs to be moved to the right-hand side of (\ref{eq:modified dynamical equation}). 
Table \ref{tab:derivatives} summarizes the unknown and known terms of the first derivatives and the product of $\Delta$ and $x$, where $\delta _{ij}$ denotes the Kronecker function. For the second derivatives of $x$, $y$, and $z$, it can be similarly summarized as shown in Table \ref{tab:second derivatives}.

Then we move all the known terms to the right-hand side of (\ref{eq:modified dynamical equation}), add them to $p_{ijkm}$, $q_{ijkm}$, and $r_{ijkm}$, and re-denote them as ${\bar p_{ijkm}}$, ${\bar q_{ijkm}}$ and ${\bar r_{ijkm}}$.
Besides, we need to address the calculation of the unknown term $\Omega_{i-1j}$ and ${\rm N}_{ij-1}$. 
In fact, they are composed of the unknown term $2\omega_{00}\omega_{i-1j}$ ($2v_{00}v_{ij-1}$) and the remaining known terms, i.e., $\Omega_{i-1j} = 2\omega_{00}\omega_{i-1j} + \sum\limits_{{i_1},{j_1},{i_2},{j_2}} {C_2^{1 - {\delta _{{i_1}{i_2}}}{\delta _{{j_1}{j_2}}}}{\omega _{{i_1}{j_1}}}{\omega _{{i_2}{j_2}}}} $, ${\rm N}_{ij-1} = 2v_{00}v_{ij-1} + \sum\limits_{{i_1},{j_1},{i_2},{j_2}} {C_2^{1 - {\delta _{{i_1}{i_2}}}{\delta _{{j_1}{j_2}}}}{v_{{i_1}{j_1}}}{v_{{i_2}{j_2}}}} $, where $i_1 + j_1 + i_2 + j_2 = n - 1$ and $C_n^k$ denotes a combination. Similar results can be obtained for the second derivatives of $y$ and $z$. In summary, the linear equation of n-order unknown coefficients is yielded
\begin{equation}
\begin{array}{l}
 - (\varpi _{km}^2 + 1 + 2{c_2}){x_{ijkm}} - 2{\varpi _{km}}{y_{xjkm}} 
  - 2\left( {{\omega _{00}} + \kappa } \right){\omega _{i - 1j}}{\delta _{1k}}{\delta _{0m}} = {{\bar {\bar p}}_{ijkm}}\\
 - 2{\varpi _{km}}{x_{ijkm}} + ({c_2} - 1 - \varpi _{km}^2){y_{xjkm}} 
 - 2\left( {\kappa {\omega _{00}} + 1} \right){\omega _{i - 1j}}{\delta _{1k}}{\delta _{0m}} = {{\bar {\bar q}}_{ijkm}}\\
({c_2} - \varpi _{km}^2){z_{ijkm}} - {d_{00} \eta}{x_{ijkm}} - 2{v_{00}}{v_{ij - 1}}{\delta _{0k}}{\delta _{1m}} 
 - 2\eta {\omega _{00}}{\omega _{i - 1j}}{\delta _{1k}}{\delta _{0m}} - {d_{i - 1j}}{\delta _{1k}}{\delta _{0m}} = {{\bar {\bar r}}_{ijkm}}
\end{array}
	\label{eq:unknown coefficients}
\end{equation}
where ${\varpi _{km}} = k{\omega _{00}} + m{v_{00}}$, ${d_{00}} = \left( {{c_2} - \omega _{00}^2} \right)\eta $, and
\begin{equation}
\begin{aligned}
{{\bar {\bar p}}_{ijkm}} &= {{\bar p}_{ijkm}} + {\delta _{k1}}{\delta _{0m}}\sum\limits_{{i_1},{j_1},{i_2},{j_2}} {C_2^{1 - {\delta _{{i_1}{i_2}}}{\delta _{{j_1}{j_2}}}}{\omega _{{i_1}{j_1}}}{\omega _{{i_2}{j_2}}}} \\
{{\bar {\bar q}}_{ijkm}} &= {{\bar q}_{ijkm}} + \kappa {\delta _{k1}}{\delta _{0m}}\sum\limits_{{i_1},{j_1},{i_2},{j_2}} {C_2^{1 - {\delta _{{i_1}{i_2}}}{\delta _{{j_1}{j_2}}}}{\omega _{{i_1}{j_1}}}{\omega _{{i_2}{j_2}}}} \\
{{\bar {\bar r}}_{ijkm}} &= {{\bar r}_{ijkm}} + \eta {\delta _{1k}}{\delta _{0m}}\sum\limits_{{i_1},{j_1},{i_2},{j_2}} {C_2^{1 - {\delta _{{i_1}{i_2}}}{\delta _{{j_1}{j_2}}}}{\omega _{{i_1}{j_1}}}{\omega _{{i_2}{j_2}}}} \\ &+ {\delta _{0k}}{\delta _{1m}}\sum\limits_{{i_1},{j_1},{i_2},{j_2}} {C_2^{1 - {\delta _{{i_1}{i_2}}}{\delta _{{j_1}{j_2}}}}{v_{{i_1}{j_1}}}{v_{{i_2}{j_2}}}} .
\end{aligned}
	\label{eq:pqr coefficients}
\end{equation}
When $(k, m) \ne (1, 0), (k, m) \ne (0, 1)$, (\ref{eq:unknown coefficients}) becomes the regular linear equations (\ref{eq:case1 pqr}). The coefficients $x_{ijkm}$, $y_{ijkm}$, and $z_{ijkm}$ can be solved immediately. 
\begin{equation}
\begin{aligned}
 - (\varpi _{km}^2 + 1 + 2{c_2}){x_{ijkm}} - 2{\varpi _{km}}{y_{xjkm}} &= {{\bar {\bar p}}_{ijkm}}\\
 - 2{\varpi _{km}}{x_{ijkm}} + ({c_2} - 1 - \varpi _{km}^2){y_{xjkm}} &= {{\bar {\bar q}}_{ijkm}}\\
({c_2} - \varpi _{km}^2){z_{ijkm}} - {d_{00}\eta}{x_{ijkm}} &= {{\bar {\bar r}}_{ijkm}}.
\end{aligned}
    \label{eq:case1 pqr}
\end{equation}
When $(k, m) = (1, 0)$, $x_{ijkm}$, $y_{ijkm}$, and $z_{ijkm}$ are couple. Thus, $x_{ijkm}$ can be set zero, $z_{ijkm} = \eta x_{ijkm} = 0$. In this case, (\ref{eq:unknown coefficients}) is turned to linear equations (\ref{eq:case2 pqr}). Then, $y_{ijkm}$ and $\omega {i-1j}$ are solved by the first two equations of (\ref{eq:case2 pqr}). $d_{i-1j}$ is obtained from ${d_{i - 1j}} =  - {\bar r_{ijkm}} - \eta \sum\limits_{{i_1},{j_1},{i_2},{j_2}} {C_2^{1 - {\delta _{{i_1}{i_2}}}{\delta _{{j_1}{j_2}}}}{\omega _{{i_1}{j_1}}}{\omega _{{i_2}{j_2}}}} $.
\begin{equation}
\begin{array}{l}
 - 2{\omega _{00}}{y_{ijkm}} - 2\left( {{\omega _{00}} + \kappa } \right){\omega _{i - 1j}} = {{\bar p}_{ijkm}} 
  + \sum\limits_{{i_1},{j_1},{i_2},{j_2}} {C_2^{1 - {\delta _{{i_1}{i_2}}}{\delta _{{j_1}{j_2}}}}{\omega _{{i_1}{j_1}}}{\omega _{{i_2}{j_2}}}} \\
 + ({c_2} - 1 - \omega _{00}^2){y_{ijkm}} - 2\left( {\kappa {\omega _{00}} + 1} \right){\omega _{i - 1j}} = {{\bar q}_{ijkm}}  + \kappa \sum\limits_{{i_1},{j_1},{i_2},{j_2}} {C_2^{1 - {\delta _{{i_1}{i_2}}}{\delta _{{j_1}{j_2}}}}{\omega _{{i_1}{j_1}}}{\omega _{{i_2}{j_2}}}} \\
 - {d_{i - 1j}} = {{\bar r}_{ijkm}}{\rm{ + }}\eta \sum\limits_{{i_1},{j_1},{i_2},{j_2}} {C_2^{1 - {\delta _{{i_1}{i_2}}}{\delta _{{j_1}{j_2}}}}{\omega _{{i_1}{j_1}}}{\omega _{{i_2}{j_2}}}} .
\end{array}
	\label{eq:case2 pqr}
\end{equation}
When $(k, m) = (0, 1)$, the coefficient of $z_{ijkm}$ is zero, and therefore $z_{ijkm}$ is set zero. In this case, (\ref{eq:unknown coefficients}) simplifies into linear equations (\ref{eq:case3 pqr}). Then, $x_{ijkm}$ and $y_{ijkm}$ are solved by the first two equations of (\ref{eq:case3 pqr}). $v_{ij-1}$ is obtained from $ - 2{v_{00}}{v_{ij - 1}} = {\bar r_{ijkm}} + {d_{00}}{x_{ijpq}} + \sum\limits_{{i_1},{j_1},{i_2},{j_2}} {C_2^{1 - {\delta _{{i_1}{i_2}}}{\delta _{{j_1}{j_2}}}}{v_{{i_1}{j_1}}}{v_{{i_2}{j_2}}}} $.
\begin{equation}
\begin{array}{l}
 - (v_{00}^2 + 1 + 2{c_2}){x_{ijkm}} - 2{v_{00}}{y_{ijkm}} = {{\bar p}_{ijkm}}\\
 - 2{v_{00}}{x_{ijkm}} + ({c_2} - 1 - v_{00}^2){y_{ijkm}} = {{\bar q}_{ijkm}}\\
 - {d_{00}}{x_{ijkm}} - 2{v_{00}}{v_{ij - 1}} = {{\bar r}_{ijkm}} + \sum\limits_{{i_1},{j_1},{i_2},{j_2}} {C_2^{1 - {\delta _{{i_1}{i_2}}}{\delta _{{j_1}{j_2}}}}{v_{{i_1}{j_1}}}{v_{{i_2}{j_2}}}} .
\end{array}
	\label{eq:case3 pqr}
\end{equation}

\section{Results}\label{result}
In this section, the third-order analytical solution for central manifolds around the collinear libration points in the RTBP with arbitrary system parameter $\mu$ is derived. Moreover, the construction of the series solution up to a certain order $n$ is implemented for the given system parameter $\mu$, such as the Sun-Earth system ($\mu$ = 3.040423398444176e-6) or Earth-Moon system ($\mu$ = 1.215058191870689e-2), utilizing the C++ 17 programming language.

\subsection{Third-order analytical solution} \label{analytical result}
It is well-known that halo/quasihalo orbits in the RTBP first appear in the third-order series solution. To achieve a comprehensive description of the center manifolds, their third-order analytical solution is derived using the analytical construction method described in Section \ref{sect:construction}, as follows:
\begin{equation*}
\begin{array}{l}
x = \alpha \cos {\theta _1} + \left( {{a_{21}} + {a_{22}}{\eta ^2}} \right){\alpha ^2} + \left( {{a_{23}} + {a_{24}}{\eta ^2}} \right){\alpha ^2}\cos 2{\theta _1}\\
 + {a_{25}}\eta \alpha \beta \cos ({\theta _1} + {\theta _2}) + {a_{26}}\eta \alpha \beta \cos ({\theta _1} - {\theta _2})\\
 + {a_{27}}{\beta ^2} + {a_{28}}{\beta ^2}\cos 2{\theta _2}\\
 + ( {{a_{31}}{\eta ^4} + {a_{32}}{\eta ^2} + {a_{33}}} ){\alpha ^3}\cos 3{\theta _1} + ( {{a_{34}}{\eta ^3} + {a_{35}}\eta } ){\alpha ^2}\beta \cos {\theta _2}\\
 + ( {{a_{36}}{\eta ^3} + {a_{37}}\eta }){\alpha ^2}\beta \cos (2{\theta _1} + {\theta _2}) \\ 
 + ( {{a_{38}}{\eta ^3} + {a_{39}}\eta } ){\alpha ^2}\beta \cos (2{\theta _1} - {\theta _2})\\
 + ( {{a_{310}}{\eta ^2} + {a_{311}}} )\alpha {\beta ^2}\cos ({\theta _1} + 2{\theta _2}) \\
 + ( {{a_{312}}{\eta ^2} + {a_{313}}})\alpha {\beta ^2}\cos ({\theta _1} - 2{\theta _2})\\
 + {a_{314}}\eta {\beta ^3}\cos {\theta _2} + {a_{315}}\eta {\beta ^3}\cos 3{\theta _2},
\\
y = \kappa \alpha \sin {\theta _1} + ( {{b_{21}} + {b_{22}}{\eta ^2}} ){\alpha ^2}\sin 2{\theta _1} + {b_{23}}\eta \alpha \beta \sin ({\theta _1} + {\theta _2})\\
 + {b_{24}}\eta \alpha \beta \sin ({\theta _1} - {\theta _2}) + {b_{25}}{\beta ^2}\sin 2{\theta _2}\\
 + ( {{b_{31}}{\eta ^4} + {b_{32}}{\eta ^2} + {b_{33}}} ){\alpha ^3}\sin {\theta _1} + ( {{b_{34}}{\eta ^4} + {b_{35}}{\eta ^2} + {b_{36}}} ){\alpha ^3}\sin 3{\theta _1}\\
 + ( {{b_{37}}{\eta ^3} + {b_{38}}\eta } ){\alpha ^2}\beta \sin {\theta _2}
 + ( {{b_{39}}{\eta ^3} + {b_{310}}\eta } ){\alpha ^2}\beta sin(2{\theta _1} + {\theta _2}) \\
 + ( {{b_{311}}{\eta ^3} + {b_{312}}\eta } ){\alpha ^2}\beta sin(2{\theta _1} - {\theta _2})
 + ( {{b_{313}}{\eta ^2} + {b_{314}}} )\alpha {\beta ^2}\sin {\theta _1}\\
 + ( {{b_{315}}{\eta ^2} + {b_{316}}} )\alpha {\beta ^2}\sin ({\theta _1} + 2{\theta _2}) \\
 + ( {{b_{317}}{\eta ^2} + {b_{318}}} )\alpha {\beta ^2}\sin ({\theta _1} - 2{\theta _2})\\
 + {b_{319}}\eta {\beta ^3}\sin {\theta _2} + {b_{320}}\eta {\beta ^3}\sin 3{\theta _2},
\end{array}
\end{equation*}

\begin{equation}
\begin{array}{l}
z = \eta \alpha \cos {\theta _1} + \left( {{d_{21}} + {d_{22}}{\eta ^2}} \right)\eta {\alpha ^2} + \left( {{d_{23}} + {d_{24}}{\eta ^2}} \right)\eta {\alpha ^2}\cos 2{\theta _1}\\
 + ( {{d_{25}} + {d_{26}}{\eta ^2}} )\alpha \beta \cos ({\theta _1} + {\theta _2}) + ( {{d_{27}} + {d_{28}}{\eta ^2}} )\alpha \beta \cos ({\theta _1} - {\theta _2})\\
 + {d_{29}}\eta {\beta ^2} + {d_{210}}\eta {\beta ^2}\cos 2{\theta _2}
 + ( {{d_{31}}{\eta ^5} + {d_{32}}{\eta ^3} + {d_{33}}\eta } ){\alpha ^3}\cos 3{\theta _1}\\
 + ( {{d_{34}}{\eta ^4} + {d_{35}}{\eta ^2} + {d_{36}}} ){\alpha ^2}\beta \cos (2{\theta _1} + {\theta _2}) \\
 + ( {{d_{37}}{\eta ^4} + {d_{38}}{\eta ^2} + {d_{39}}} ){\alpha ^2}\beta \cos (2{\theta _1} - {\theta _2})\\
 + \left( {{d_{310}}{\eta ^3} + {d_{311}}\eta } \right)\alpha {\beta ^2}\cos ({\theta _1} + 2{\theta _2}) \\
 + ( {{d_{312}}{\eta ^3} + {d_{313}}\eta } )\alpha {\beta ^2}\cos ({\theta _1} - 2{\theta _2})\\
 + ( {{d_{314}}{\eta ^2} + {d_{315}}} ){\beta ^3}\cos 3{\theta _2},
\end{array}
	\label{eq:3-order solution}
\end{equation}
with the frequencies 
\begin{equation}
\begin{array}{l}
\omega  = {\omega _0} + {\omega _{20}} + {\omega _{02}} 
\\ = {e_{31}}{\alpha ^2}{\eta ^4} + \left( {{e_{32}}{\alpha ^2} + {e_{33}}{\beta ^2}} \right){\eta ^2} + {e_{34}}{\alpha ^2} + {e_{35}}{\beta ^2} + {\omega _0}\\
v = {v_0} + {v_{20}} + {v_{02}} 
\\ = {e_{36}}{\alpha ^2}{\eta ^4} + \left( {{e_{37}}{\alpha ^2} + {e_{38}}{\beta ^2}} \right){\eta ^2} + {e_{39}}{\alpha ^2} + {e_{310}}{\beta ^2} + {v_0},
\end{array}
	\label{eq:3-order frequencies}
\end{equation}
and the coupling correction term
\begin{equation}
\begin{array}{l}
\eta \Delta  = \eta \left( {{d_{00}} + {d_{20}}{\alpha ^2} + {d_{02}}{\beta ^2}} \right) 
\\ = \eta \left[ {{l_1}{\eta ^4}{\alpha ^2} + \left( {{l_2}{\alpha ^2} + {l_3}{\beta ^2}} \right){\eta ^2} + {l_4}{\alpha ^2} + {l_5}{\beta ^2} - \left( {\omega _0^2 - v_0^2} \right)} \right] = 0
\end{array}
	\label{eq:3-order coupling}
\end{equation}
where $a_{ij}$, $b_{ij}$, $d_{ij}$, $e_{ij}$, and $l_i$ are constant as provided in \ref{A:analytical Coefficients}.

\begin{figure}
    \centering
	\includegraphics[width=0.5\columnwidth]{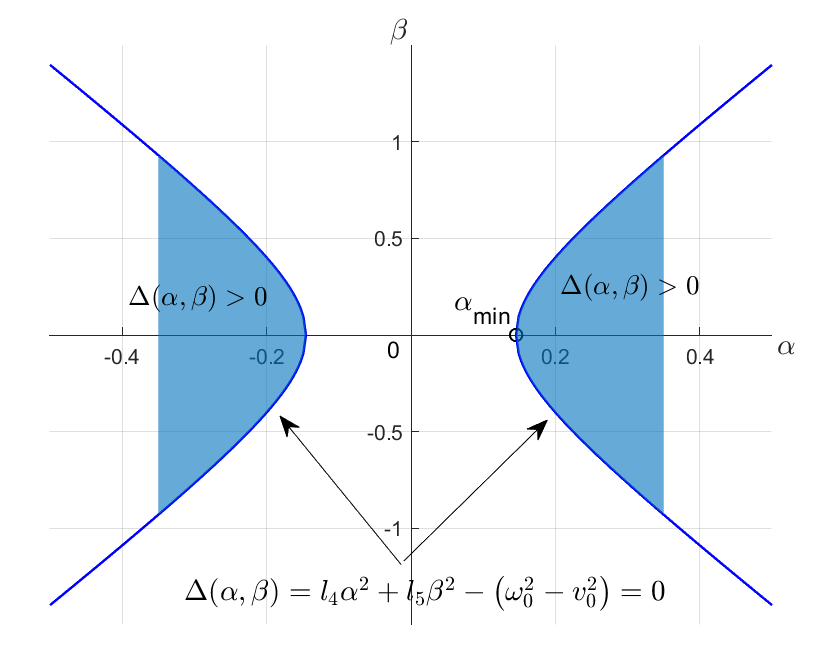}
    \caption{Hyperbola for the bifurcation equation (\ref{eq:3-order bifurcation}) with $\alpha$ and $\beta$.}
    \label{fig:Fig Hyperbola}
\end{figure}

Equation (\ref{eq:3-order coupling}) establishes an explicit relationship between $\eta$ and $\alpha$ and $\beta$, i.e., $\eta = \eta(\alpha, \beta)$. It obviously has a trivial solution $\eta = 0$ for any values of $\alpha$ and $\beta$. In this case, the third-order solution (\ref{eq:3-order solution}) describes Lissajous orbits. With the increment of $\alpha$ and $\beta$, other non-zero real solutions will bifurcate if we have
\begin{equation}
    \Delta \left( {\alpha ,\beta ,\eta } \right) = {l_1}{\eta ^4}{\alpha ^2} + ( {{l_2}{\alpha ^2} + {l_3}{\beta ^2}} ){\eta ^2} + {l_4}{\alpha ^2} + {l_5}{\beta ^2} - ( {\omega _0^2 - v_0^2} ) = 0.
	\label{eq:3-order bifurcation}
\end{equation}
In this case, solution (\ref{eq:3-order solution}) describes quasihalo orbits. To find the critical condition for bifurcation, we let $\eta = 0$ in (\ref{eq:3-order bifurcation}) and obtain 
\begin{equation}
\Delta \left( {\alpha ,\beta } \right) = {l_4}{\alpha ^2} + {l_5}{\beta ^2} - ( {\omega _0^2 - v_0^2} ) = 0
	\label{eq:3-order critical condition}
\end{equation}
It is not difficult to calculate and verify that $l_1 > 0, l_2 < 0, l_3 < 0, l_4 > 0$ and $l_5 < 0$ for the three collinear libration points with all system parameter $\mu  \in \left( {0,0.5} \right]$. Hence, (\ref{eq:3-order critical condition}) is a hyperbolic equation where bifurcation occurs and suitable non-zero real solutions exist for (\ref{eq:3-order bifurcation}) when $\Delta (\alpha, \beta)$ as illustrated in Fig. \ref{fig:Fig Hyperbola}. 
Particularly, if the amplitude $\beta$ is set to zero, the third-order solution (\ref{eq:3-order solution}) describes halo orbits. It is seen from (\ref{eq:3-order critical condition}) that the minimum permissible value of $\alpha$ is given as (Without loss of generality, only the case of $\alpha > 0 $ is considered in the following)
\begin{equation}
{\alpha _{\min }} = \sqrt {\left( {\omega _0^2 - v_0^2} \right)/{l_4}} .
	\label{eq:3-order min alpha}
\end{equation}
When $\alpha > \alpha _{\min }$ and $\beta = 0$, by solving (\ref{eq:3-order bifurcation}) we can find two solutions of $\eta$,
\begin{equation}
{\eta ^2} = \frac{{ - {l_2}{\alpha ^2} - \sqrt {{l_2}^2{\alpha ^4} - 4{l_1}{\alpha ^2}\left( {{l_4}{\alpha ^2} - \omega _0^2 + v_0^2} \right)} }}{{2{l_1}{\alpha ^2}}} > 0
	\label{eq:3-order eata}
\end{equation}
The solution branch corresponds to northern halo orbits (Class I) for $\eta > 0$ and southern halo orbits (Class II) for $\eta < 0$. These results align with the classical outcomes of the third-order analytical solution of halo orbits presented in \cite{richardson1980analytic}. 
To sum up, quasihalo orbits bifurcate from Lissajous orbits when $\eta$ is a solution of (\ref{eq:3-order bifurcation}). In particular, halo orbits bifurcate from planar Lyapunov periodic orbits when $\eta$ is a solution of (\ref{eq:3-order bifurcation}) with $\beta$ = 0.

\textit{Remark 4}. The third-order solution (\ref{eq:3-order solution}) serves as an initial approximation to the center manifolds of around the collinear libration points in the RTBP. A higher-order series solution is required for real space missions. In such case, the bifurcation equation is no longer a hyperbola. Figure \ref{fig:Fig Hyperbola} just provides a basic outline of the feasible region for $\eta$, and in fact, its real feasible region is more intricate than depicted in Fig. \!\ref{fig:Fig Hyperbola}. A higher-order numerical feasible region of $\eta$ will be presented in the following section.

\begin{figure}
    \centering
	\includegraphics[width=1\columnwidth]{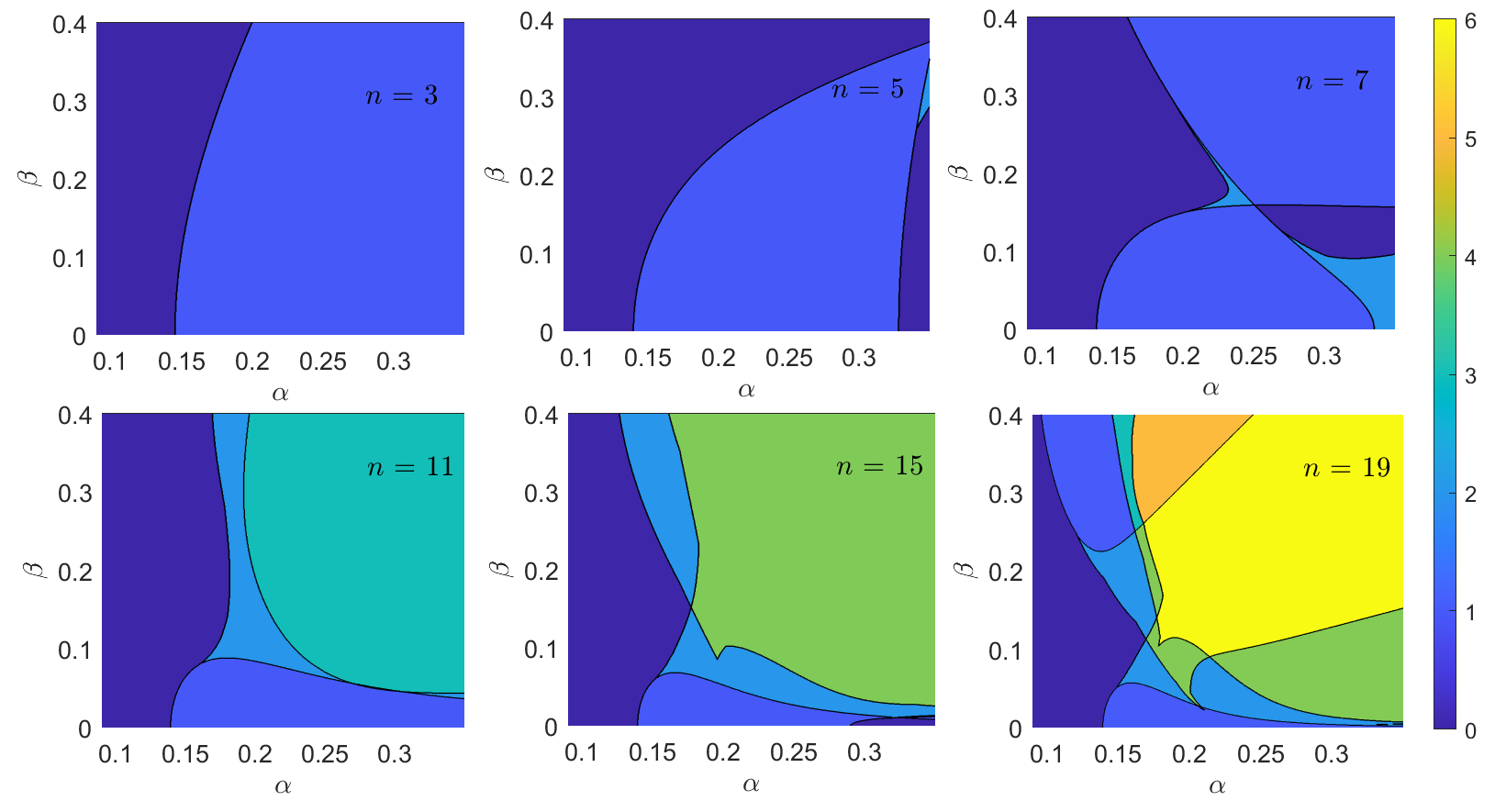}
    \caption{Feasible region of coupling coefficient $\eta$ for center manifolds of $L_1$ in the Earth-Sun system with different orders $n$ = 3, 5, 7, 11, 15, 19. (In this case, $\mu$ = 3.040423398444176e-6, $\gamma_1$ = 1.00109772277814e-2) with $\alpha$ and $\beta$.}
    \label{fig:Feasible region }
\end{figure}
\begin{figure}
\centering
	\includegraphics[width=1\columnwidth]{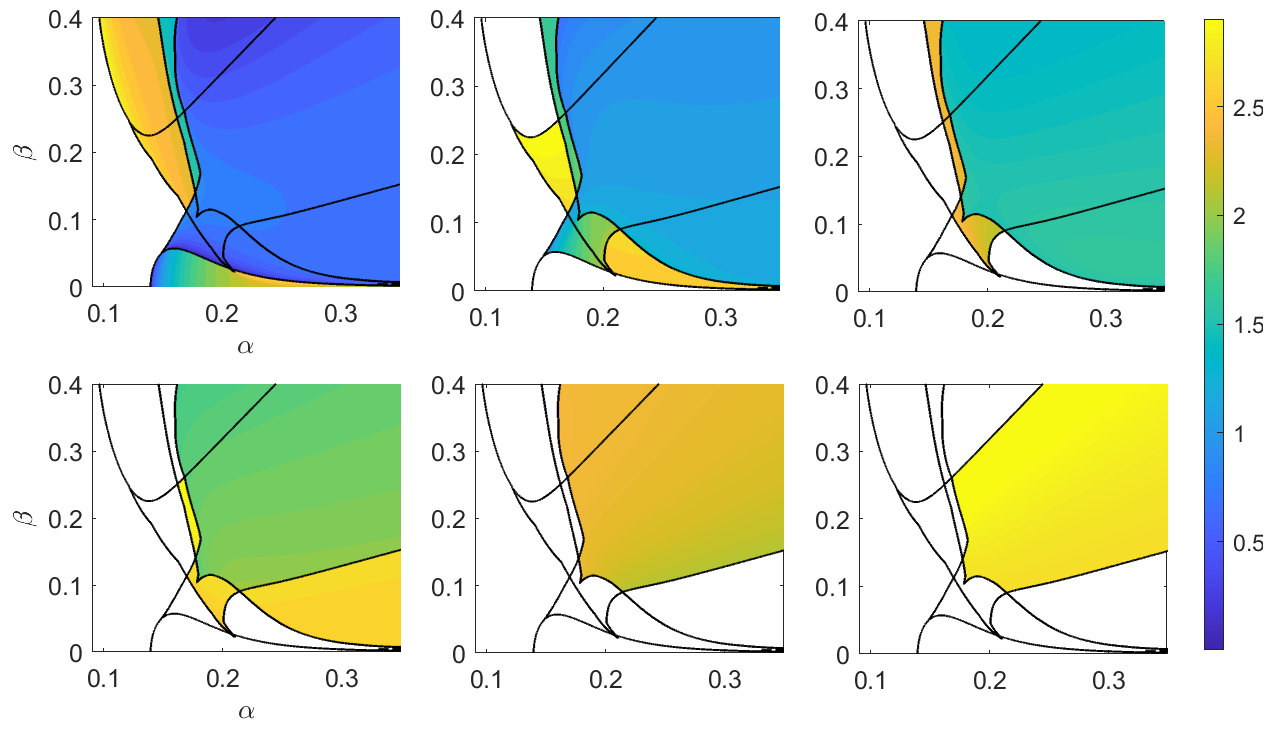}
    \caption{Contour maps of coupling coefficient $\eta$ within feasible region for center manifolds of $L_1$ in the Earth-Sun.}
    \label{fig:coupling coefficient }
\end{figure}

\subsection{Numerical results}\label{numerical result}
The semi-analytical computation of center manifolds up to a certain order $n$ for the given system parameter $\mu$ is implemented to verify the accuracy of the proposed method. \ref{B:numerical Coefficients} shows the coefficients of expansion for center manifolds of $L_1$ in Sun-Earth system, up to order 3.

The third-order analytical solution in Subsection \ref{analytical result} illustrates that a feasible region of coupling coefficient $\eta$ of center manifolds is bounded by a hyperbola. Similar to the process in (\ref{eq:3-order critical condition}) for the third-order solution, higher-order $\Delta (\alpha, \beta)$ can be obtained during the computation of higher-order semi-analytical solutions. For each given paired amplitudes $(\alpha, \beta)$, non-trivial values of $\eta$ can be determined from $\Delta (\alpha, \beta) = 0$. Then, The feasible region of $\eta$ of center manifolds around $L_1$ in the Earth-Sun system with different orders is presented in Fig. 2, where only the range of $\alpha \in [0, 0.35]$, $\beta \in [0, 0.4]$ and $\eta \in (0, 3.0]$ is considered due to the symmetry of $\eta$ and the divergence of (17) for large amplitudes. The colorbar represents the number of solutions $N (\eta)$ of $\Delta (\alpha, \beta) = 0$ with the range $\eta \in (0, 3.0]$. It can be seen that $N$ has only two possible values $N = 0$ or $N = 1$, distributed on both sides of a hyperbola boundary when the order $n = 3$. This result is consistent with the analytical solution in Subsection \ref{analytical result}. As the order increases, parts of the region with $N = 0$ and $N = 1$ are replaced by the region with $N(\eta) > 1$. This means that, with increasing amplitudes $\alpha$ and $\beta$, more than one quasihalo orbit bifurcates from a Lissajous orbit. 

\begin{figure}
\centering
	\includegraphics[width=0.6\columnwidth]{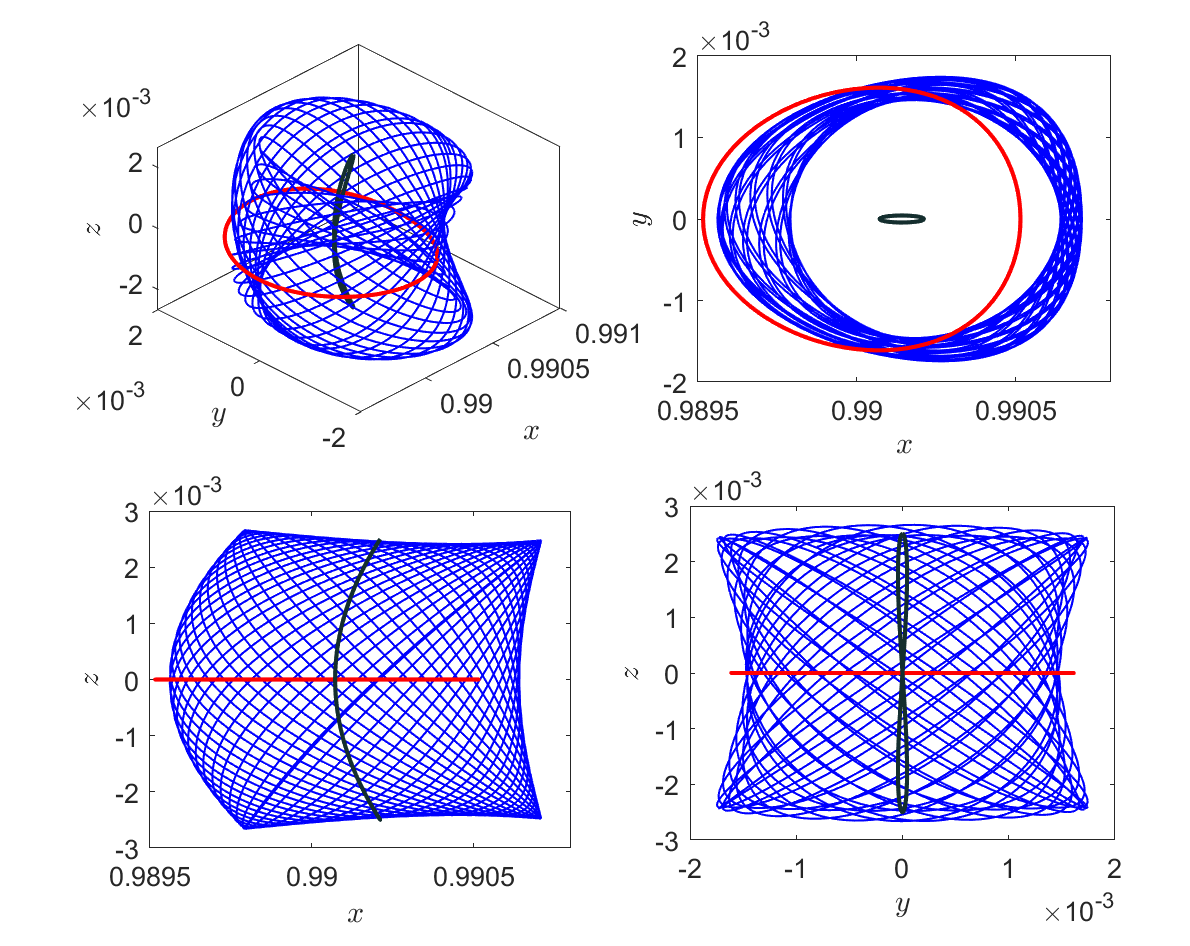}
    \caption{Periodic and quasi-periodic orbits around $L_1$ in the Earth-Sun system and its projection on $xy$, $xz$, and $yz$ planes, red: planar Lyapunov orbit with $\alpha$ = 0.05, $\beta$ = 0.0; black: vertical Lyapunov orbit with $\alpha$ = 0.0, $\beta$ = 0.25; blue: Lissajous orbit with $\alpha$ = 0.05, $\beta$ = 0.25.}
    \label{fig:Lissajous orbit }
\end{figure}

\begin{figure}[hp]
\begin{center}
\subfigure[]{
	\includegraphics[width=0.48\columnwidth]{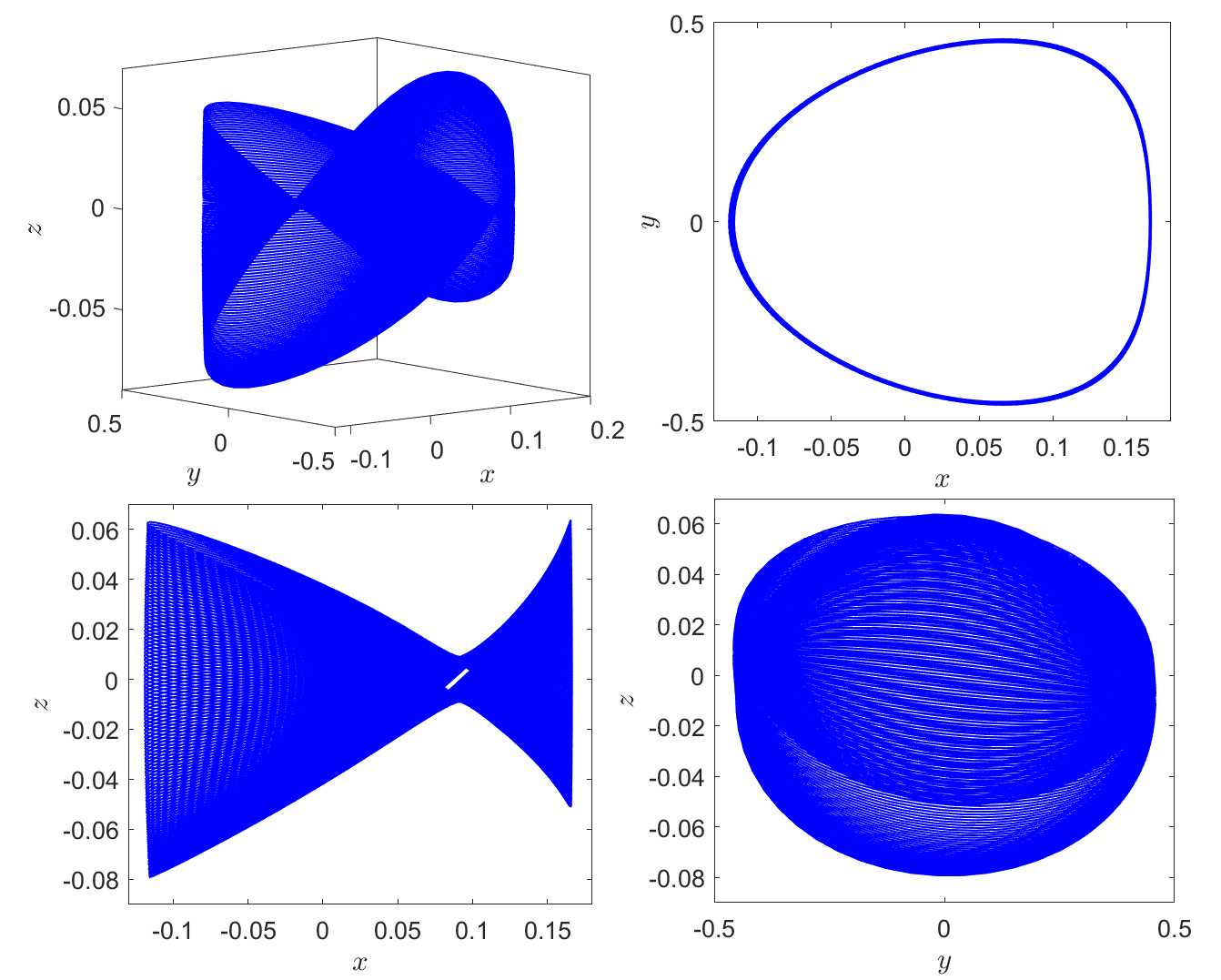}
}
\subfigure[]{
	\includegraphics[width=0.48\columnwidth]{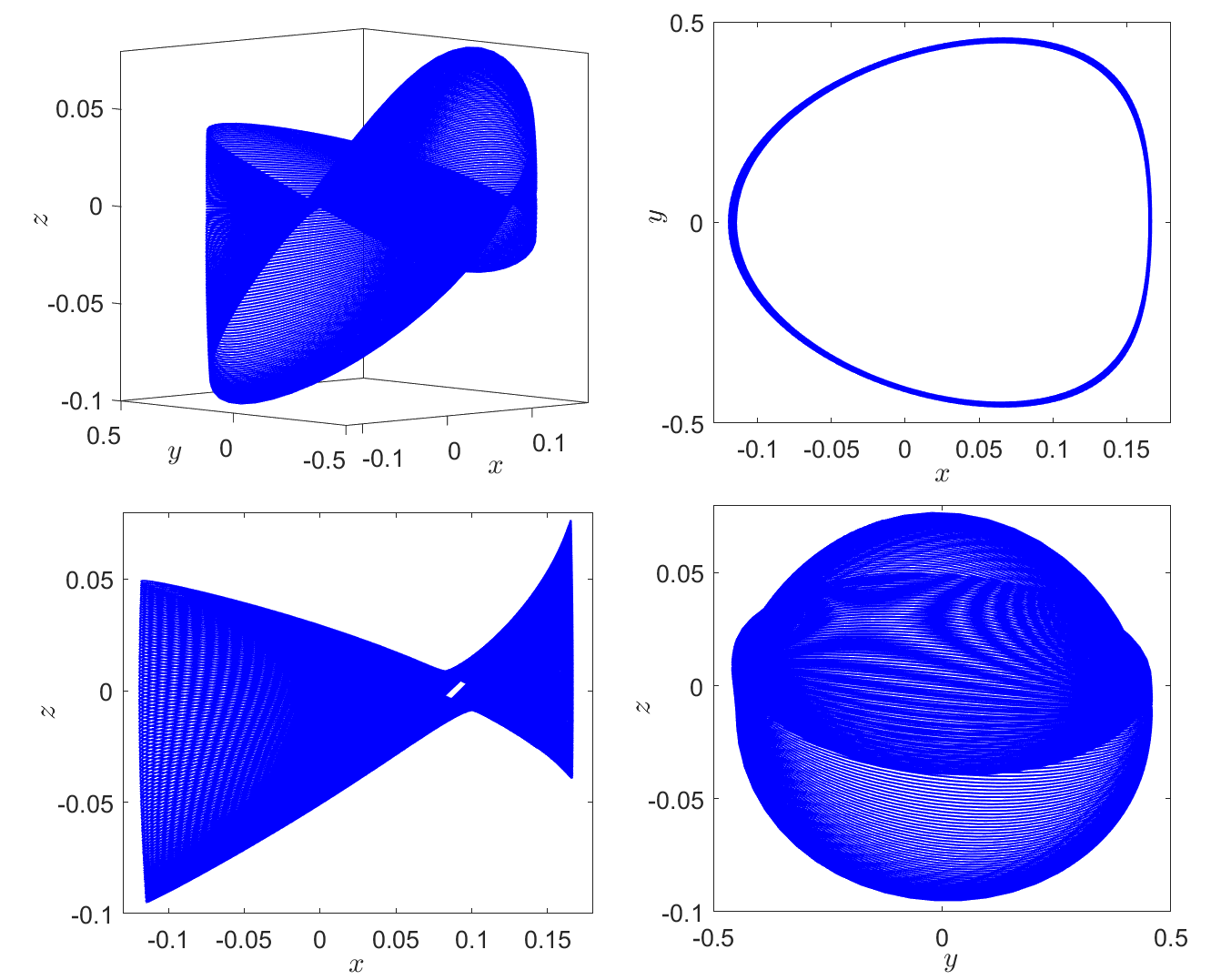}
}
    \caption{Quasi-periodic orbits around $L_1$ in the Earth-Sun system with $\alpha$ = 0.144227 and $\beta$ = 0.04 and its projection on $xy$, $xz$, and $yz$ planes. (a) $\eta$ = 0, Lissajous orbit (b) $\eta$ = 0.08180669069, quasihalo orbit.}
    \label{fig:Lissajous and quasihalo orbit}
\end{center}
\end{figure}

\textit{Remark 5}. The polynomials $\Delta (\alpha, \beta)$ may encompass all the information about local bifurcation around collinear libration points in the RTBP. It is reasonable to believe that with further increases in order, the number of solutions of $\eta$ will also increase. Furthermore, it is observed that the number of solutions for $\eta$ multiplies (due to only $\eta \in (0, 3.0]$ being considered, some solutions for $\eta$ are omitted here). This is a typical period-doubling bifurcation, indicating that chaos naturally occurs in the center manifolds of collinear libration points in the RTBP. However, the high-order series solution (\ref{eq:high-order solution}) cannot describe this phenomenon due to its divergence. 

Then, a contour map of $\eta$ values with an order $n = 19$ is computed as shown in Fig. \ref{fig:coupling coefficient }. The feasible region corresponds to the bottom-right part of Fig. \ref{fig:Feasible region }. Each contour map describes the distribution of one solution of $\eta$. Empty regions indicate no solution for $\eta$. Six contour maps represent a maximum of six solutions of $\eta$ at the 19-th order. Figure \ref{fig:coupling coefficient } shows that the feasible region is largest when there is only one non-zero solution for $\eta$ (top left). Subsequently, as the number of non-zero $\eta$ solution increases, the feasible region becomes small. When the amplitude $\alpha$ and $\beta$ are small, there are no non-zero $\eta$, corresponding to the empty region in the top-left part of Fig. \ref{fig:coupling coefficient }. In this case, the higher-order series (\ref{eq:high-order solution}) only describes planar and vertical Lyapunov orbits and Lissajous orbits. Figure \ref{fig:Lissajous orbit } shows a typical plot of planar Lyapunov orbits, vertical Lyapunov orbits and Lissajous orbits in the synodic coordinate system. 

Upon increasing $\alpha$ and $\beta$ to the region with one solution of $\eta$, Fig. \ref{fig:coupling coefficient } shows that $\eta$ values change from zero to non-zeros but they are very small, which means the week coupling of two degrees of freedom (in-plane and out-of-plane motions). In this case, quasihalo orbits closely resemble the corresponding Lissajous orbits, with the amplitudes identical to quasihalo orbits but $\eta$ values being zeros. This observation is evident from Fig. \ref{fig:Lissajous and quasihalo orbit}, where $\eta$ = 0.08180669069. As the coupling effect ($\eta$ > 0) of the in-plane motion acts on the out-of-plane motion, a plane-symmetric Lissajous orbit (linear part in $z$-direction: $z_1 = \beta \cos \theta_2$) becomes an approximately oblique-upward symmetric quasihalo orbit (linear part in $z$-direction: $z_1 = \beta \cos \theta_2  + \eta \alpha \cos \theta_1$).

With further increment in $\alpha$ and $\beta$, one $\eta$ value rapidly grows larger and more solutions for $\eta$ appear. Figure \ref{fig:Two quasihalo orbits} shows two quasihalo orbits with the same amplitudes. It can be seen that the quasihalo orbit with the smaller value of $\eta$ exhibits week coupling and remains similar to Lissajous orbits. The larger value of $\eta$ leads to a significant coupling effect on the motion in z-directions, resulting in a typical quasihalo orbit, as shown in Fig. \ref{fig:Two quasihalo orbits}b. In this case, in-plane and out-of-plane motions are strong coupling. As $\alpha$ and $\beta$ increase to the region shown in the bottom-left of Fig.\ref{fig:coupling coefficient }, four solutions for $\eta$ can be found from $\Delta = 0$, i.e., $N (\eta) = 4$. Figure \ref{fig:Four quasihalo orbits} shows four quasihalo orbits corresponding to these four solutions for the given $\alpha$ and $\beta$. It can be seen they have some strange bendings, likely due to the corresponding $\alpha$ and $\beta$ not being within the practical region of convergence.

Now, we analyze the practical convergence domain of the proposed analytical solution by comparing it with numerical solutions. Firstly, an initial condition is obtained for given amplitudes $\alpha$ and $\beta$ from the analytical solution (\ref{eq:high-order solution}). Subsequently, numerical integration of the dynamical equations is performed over a normalized time length T = $\pi$. The accuracy of the analytical solution is determined by comparing the Euclidean norm of difference in position vectors at final time between the analytical and numerically integrated solutions. Performing the same procedure for each pair of amplitudes ($\alpha$, $\beta$) within a given range yields the practical convergence domain of the solution with a specified order. 

\begin{figure}
\begin{center}
\subfigure[]{
	\includegraphics[width=0.48\columnwidth]{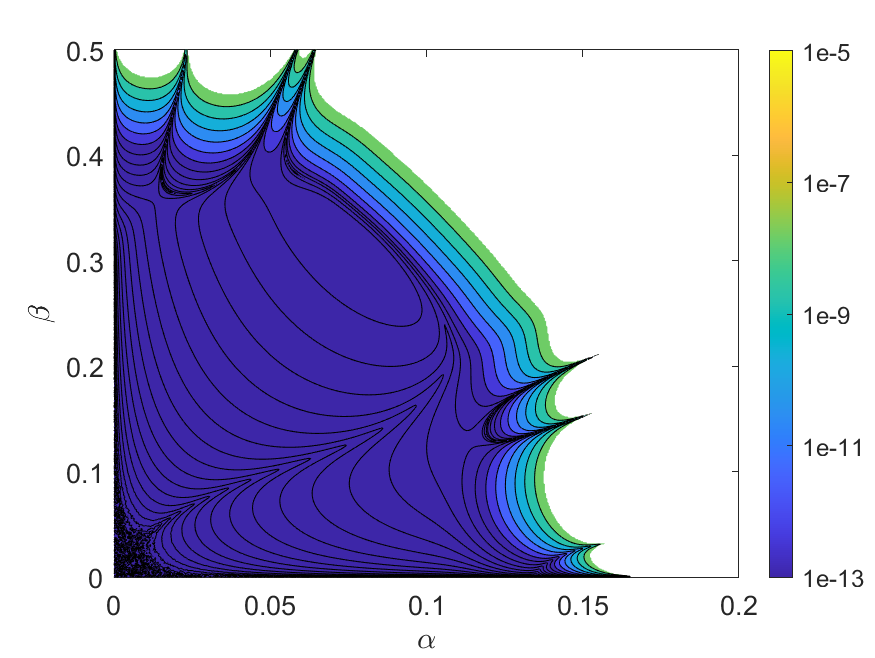}
}
\subfigure[]{
	\includegraphics[width=0.48\columnwidth]{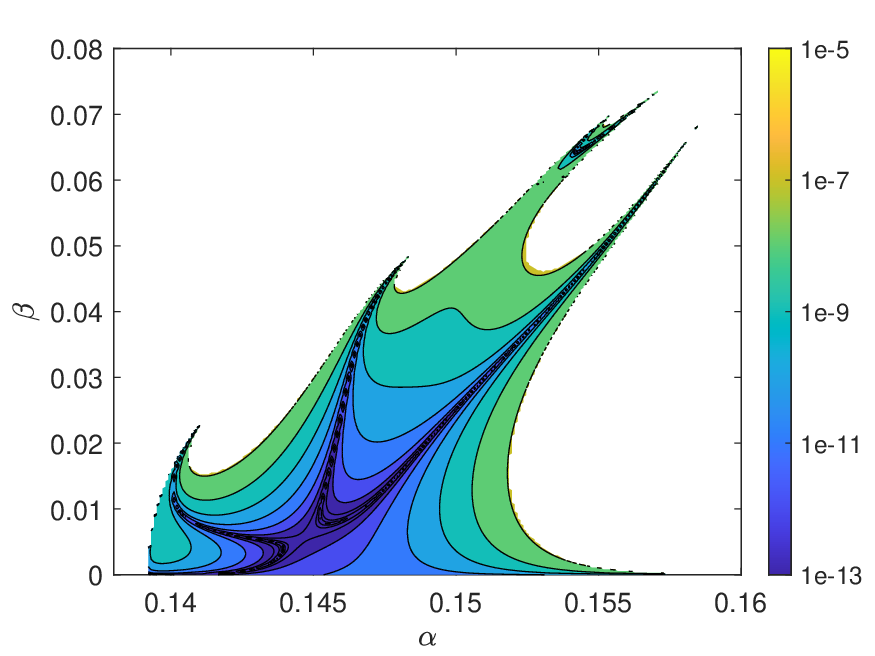}
}
    \caption{Domain of practical convergence of the proposed analytical solution up to order $n$ = 35 for (a) Lissajous orbits and (b) Quasihalo orbits around $L_1$ of the Earth-Sun system}
    \label{fig:Convergence domain}
\end{center}
\end{figure}

\begin{figure}
\begin{center}
\subfigure[]{
	\includegraphics[width=0.48\columnwidth]{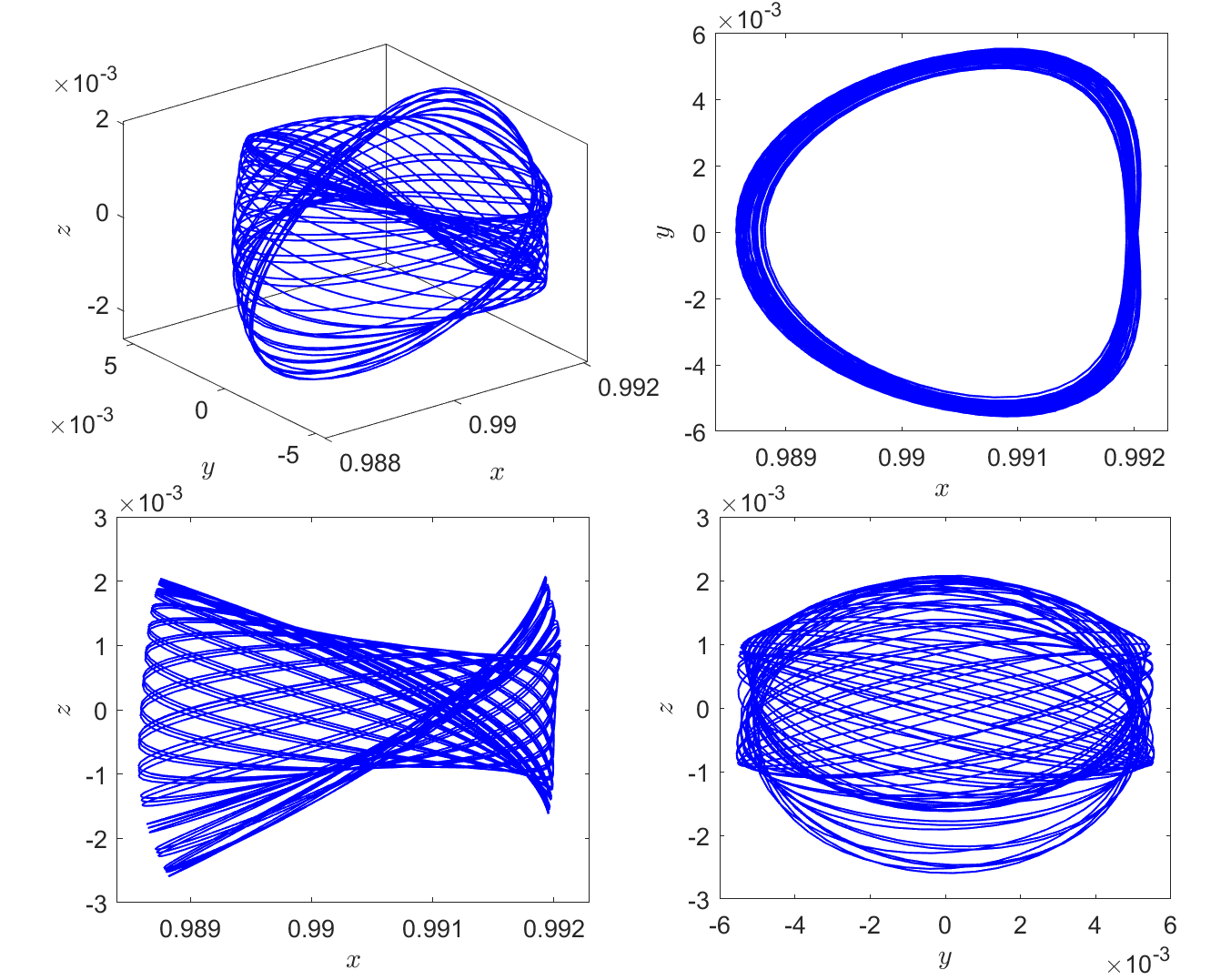}
}
\subfigure[]{
	\includegraphics[width=0.48\columnwidth]{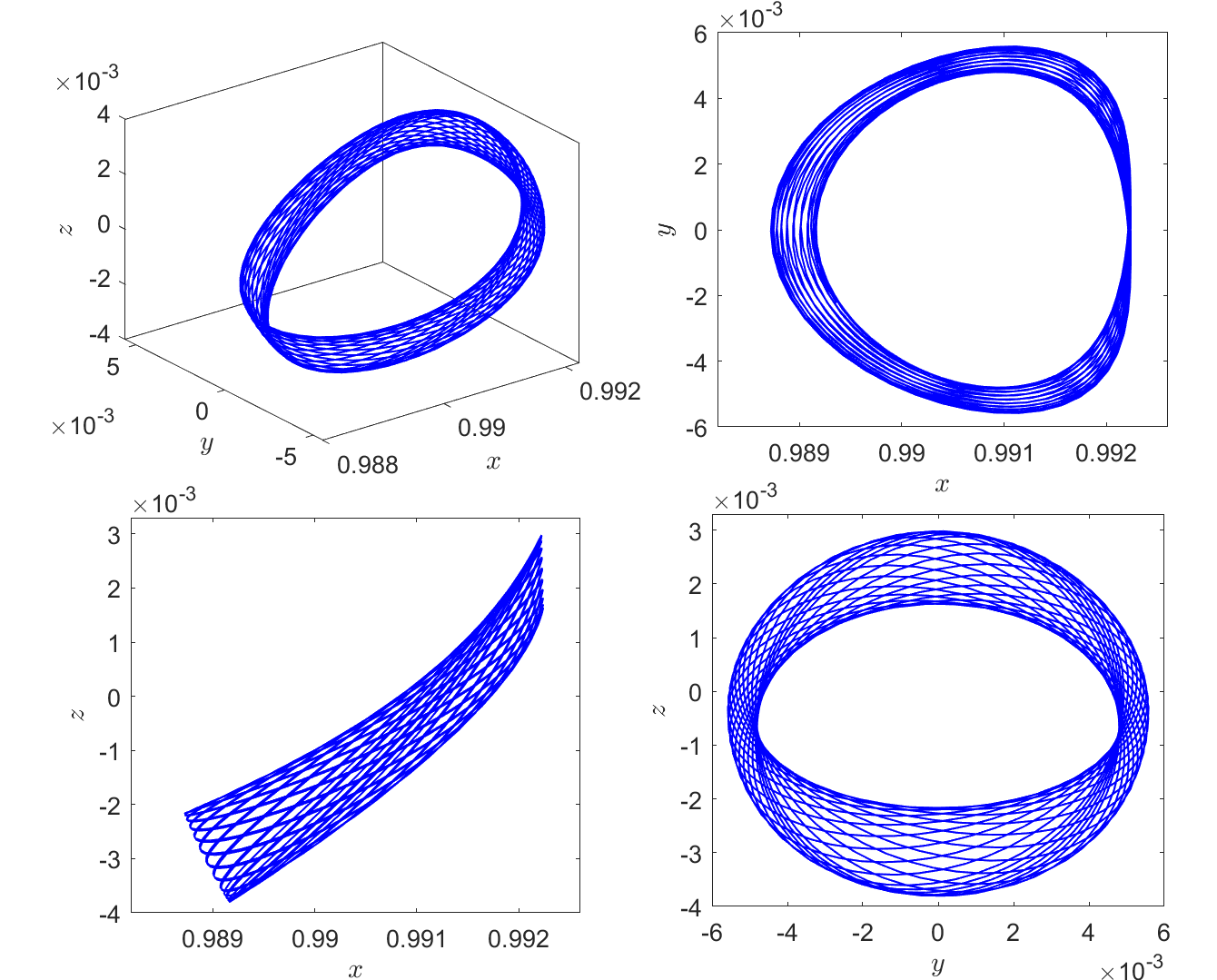}
}
    \caption{Two quasihalo orbits around $L_1$ in the Earth-Sun system with $\alpha$ = 0.167 and $\beta$ = 0.055 and its projection on $xy$, $xz$, and $yz$ planes. (a) $\eta$ = 0.04676813553 (b) $\eta$ = 1.552696086.}
    \label{fig:Two quasihalo orbits}
\end{center}
\end{figure}

\begin{figure}
\begin{center}
\subfigure[]{
	\includegraphics[width=0.48\columnwidth]{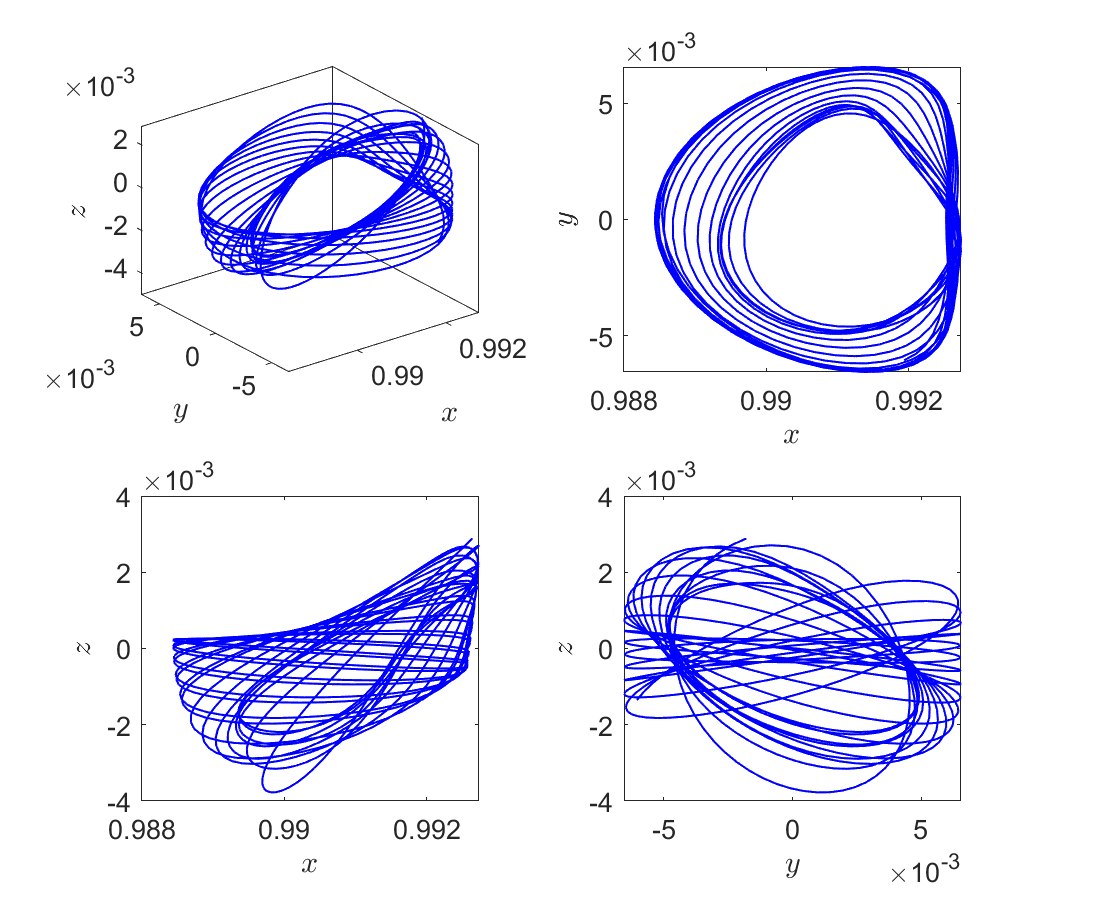}
}
\subfigure[]{
	\includegraphics[width=0.48\columnwidth]{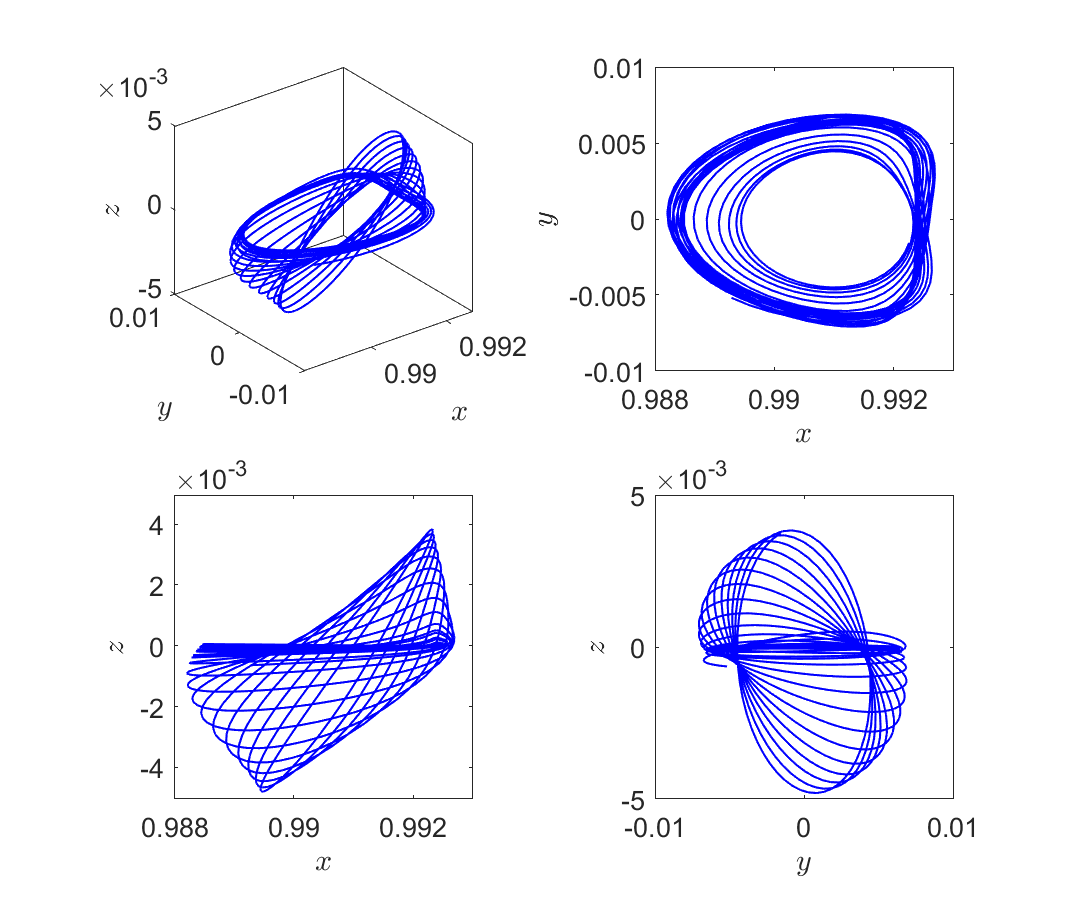}
}
\subfigure[]{
	\includegraphics[width=0.48\columnwidth]{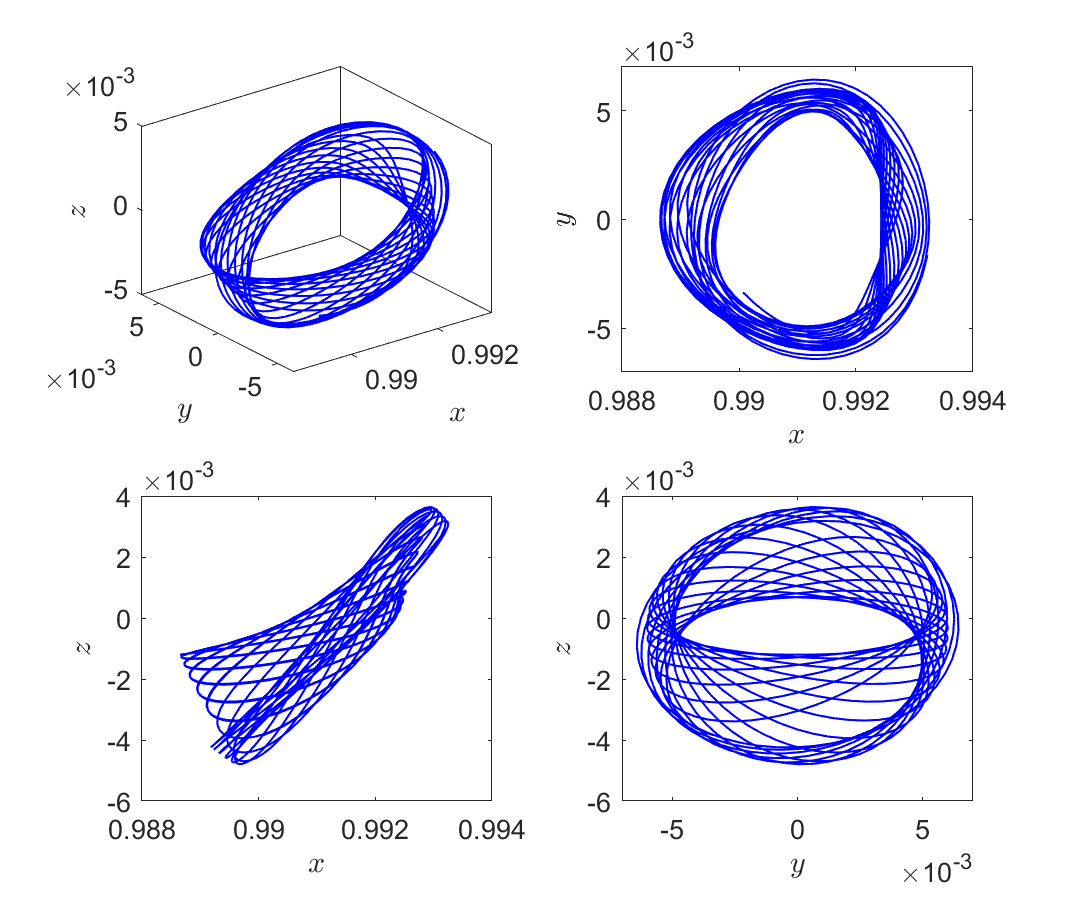}
}
\subfigure[]{
	\includegraphics[width=0.48\columnwidth]{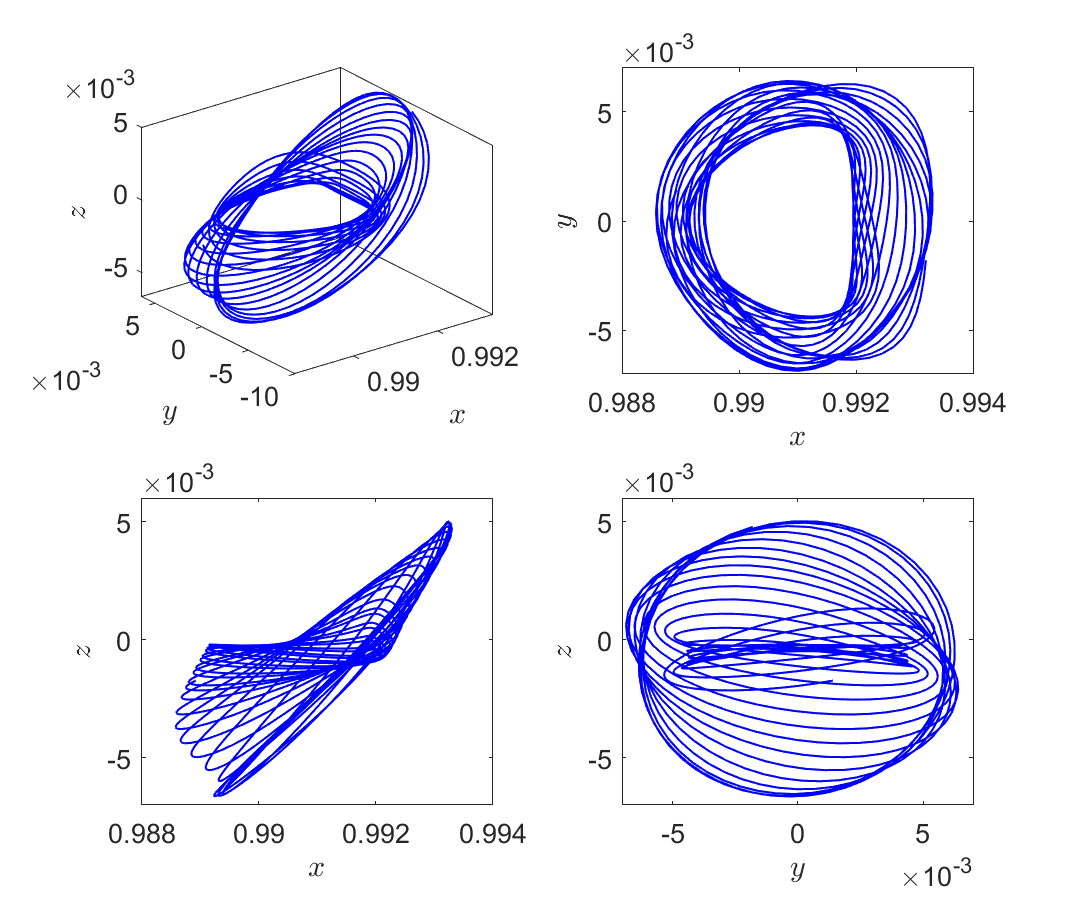}
}
    \caption{Four quasihalo orbits around $L_1$ in the Earth-Sun system with $\alpha$ = 0.180 and $\beta$ = 0.142 and its projection on $xy$, $xz$, and $yz$ planes. (a) $\eta$ = 0.9190215098 (b) $\eta$ = 1.12844267 (c) $\eta$ = 1.636802107 (d) $\eta$ = 1.870762835.}
    \label{fig:Four quasihalo orbits}
\end{center}
\end{figure}

Figure \ref{fig:Convergence domain} shows the domain of practical convergence of the proposed analytical solution up to order 35 for Lissajous orbits ($\eta = 0 $) and Quasihalo orbits ($\eta \neq 0 $) around $L_1$ of the Earth-Sun system. Here Fig. \ref{fig:Convergence domain}(a) is similar the result in \cite{jorba1999dynamics}. This is because, when $\eta$ is zero, no bifurcation occurs, and the analytical solution is identical to the solution of Lissajous orbits presented in \cite{jorba1999dynamics}. However, when $\eta$ is non-zero, coupling effects between different degrees of freedom lead to the bifurcation, specifically the generation of quasihalo orbits. Figure \ref{fig:Convergence domain}(b), for the first time, provides the actual convergence domain of the approximate analytical solution for quasihalo orbits generated by the first bifurcation from Lissajous orbits. It can be observed that the practical convergence domain for quasihalo orbits is significantly smaller than the practical convergence domain for the Lissajous orbit. This is due to the coupling in the plane direction causing a large actual amplitude in the z-direction, even when the amplitude $\beta$ is small.

It is known that center manifolds around collinear libration points in the RTBP are four-dimensional, making direct graphical display challenging. As a byproduct of the proposed analytical method for constructing center manifolds, the dynamics inside center manifolds can be globally described in a two-dimensional Poincaré section by fixing $z = 0$ with $z > 0$ and the Jacobian integral $C = C_0$ in synodic coordinate system. First, arbitrary phase angles $\phi_1$ and $\phi_2$ and arbitrary initial time t0 from (\ref{eq:high-order solution}) is chosen. For the chosen $C_0$, paired amplitudes ($\alpha$, $\beta$) and corresponding initial states ($x_0 y_0 z_0$) can be computed. Then, starting from each initial state, every points $(x_i, y_i)$ is plotted in the two-dimensional Poincaré section when $z_i = 0$ within a specified time interval according to (\ref{eq:high-order solution}).

\begin{figure}
\begin{center}
\subfigure[]{
	\includegraphics[width=0.48\columnwidth]{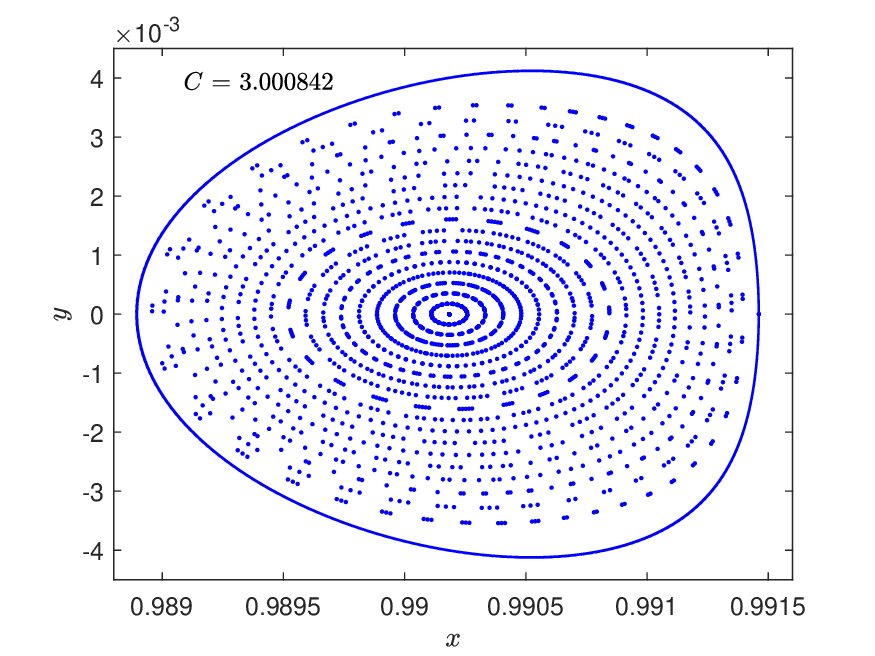}
}
\subfigure[]{
	\includegraphics[width=0.48\columnwidth]{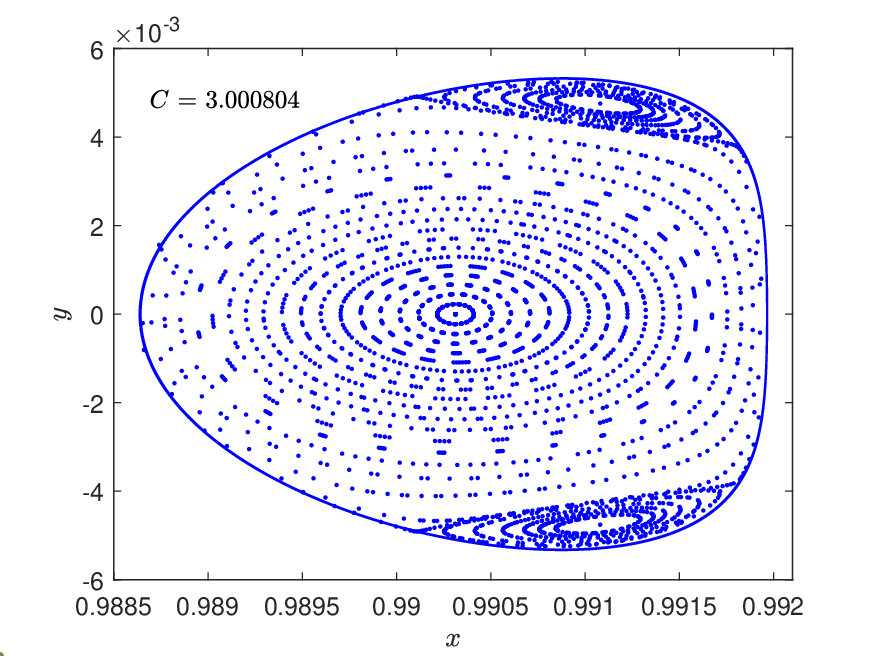}
}
    \caption{Poincaré section of the center manifold of $L_1$ in the Earth-Sun system: (a) $C$ = 3.000842. (b) $C$ = 3.000804.}
    \label{fig:Poincare section}
\end{center}
\end{figure}

Figure \ref{fig:Poincare section} shows two Poincaré sections with two different Jacobian integrals, defining a closed region. The boundary of the region is a planar Lyapunov orbit with the paired amplitudes ($\alpha_{\rm {max}}$, 0) and a fixed point on $y$-axis represents a vertical Lyapunov orbit with the paired amplitudes (0, $\beta_{\rm {max}}$), where $\alpha_{\rm{max}}$ and $\beta_{ \rm{max}}$ are maximum in-plane and out-of-plane amplitudes for the selected $C_0$, respectively. 
Other circles inside this region respond to Lissajous orbits with the paired amplitudes ($\alpha$, $\beta$) where $\alpha$ < $\alpha_{\rm{max}}$ and $\beta < \beta_{\rm{max}}$. The bifurcation occurs when the increase in $C$ results in a non-zero solution $\eta \ne 0$ for the equation $\Delta  = \sum\limits_{0 \le i+j \le n} {{d_{ij}}{\alpha ^i}{\beta ^j} = 0}$. This visual computing not only shows the emergence of well-known halo orbits from planar Lyapunov orbits but also illustrates the generation of quasi-periodic orbits from Lissajous orbits.

\textit{Remark 6}. The computed results presented in Fig. \ref{fig:Poincare section} are similar to those found in previous studies \citep{gomez2001dynamics,jorba1999dynamics}. However, these results were computed through numerical continue methods with high computational complexity or displayed in normal coordinate via a complicated normal form. In contrast, the Poincaré sections of center manifolds can be directly derived from the high-order series solution (\ref{eq:high-order solution}) in this paper.

\section{Conclusions}

Understanding the coupling between in-plane and out-of-plane motions and its connection to the bifurcation of halo/quasihalo orbits is crucial for unraveling the mechanism governing their generation and obtaining their unified an analytical solution. This paper introduced a novel concept of coupling coefficients into the RTBP for the first time, incorporated a new correction term into the RTBP equation to characterize this coupling effect and successfully derive a unified analytical solution for the center manifolds of collinear libration points in the RTBP. When the bifurcation equation $\Delta = 0$ has no real solutions, the series solution (\ref{eq:high-order solution}) describes planar Lyapunov orbits, vertical Lyapunov orbits and Lissajous orbits. When the equation $\Delta = 0$ has only zero solutions, the bifurcation occurs. When the equation $\Delta = 0$ yields non-zero real solution, the series solution (\ref{eq:high-order solution}) further describes both halo orbits and quasihalo orbits, with the latter being constructed analytically for the first time. Remarkably, larger amplitudes lead to multiple solutions of the equation $\Delta = 0$ with the higher-order series solution, indicating the discovery of a multiple bifurcation, which can be explicitly calculated.

Although the proposed analytical method is aimed at constructing center manifolds around collinear libration points in the RTBP, it can be extended to invariant manifolds in the RTBP and even general multi-degree-of-freedom dynamical systems. Besides, the coupling efficient $\eta$ and the equation $\Delta = 0$ serve as conditions for bifurcation. Thus, this method can also be referred to as an analytical bifurcation calculation method.

\appendix

\section{Coefficients of third-order analytical solution} \label{A:analytical Coefficients}
\begin{flalign*}
& {a_{21}} = \frac{{3{c_3}({\kappa ^2} - 2)}}{{4(1 + 2{c_2})}},
{a_{22}} = \frac{{3{c_3}}}{{4(1 + 2{c_2})}},
{a_{23}} =  - \frac{{3{c_3}}}{4}\frac{{({\kappa ^2} + 2)(\varpi _{20}^2 - {c_2} + 1) + 4\kappa {\varpi _{20}}}}{{{s_{20}}}}\\
& {a_{24}} = \frac{{3{c_3}}}{4}\frac{{(\varpi _{20}^2 - {c_2} + 1)}}{{{s_{20}}}},
{a_{25}} = \frac{{3{c_3}}}{2}\frac{{\varpi _{11}^2 - {c_2} + 1}}{{{s_{11}}}},
{a_{26}} = \frac{{3{c_3}}}{2}\frac{{\varpi _{1 - 1}^2 - {c_2} + 1}}{{{s_{1 - 1}}}}\\
& {a_{27}} = \frac{{3{c_3}}}{{4(1 + 2{c_2})}},
{a_{28}} = \frac{{3{c_3}}}{4}\frac{{\varpi _{02}^2 - {c_2} + 1}}{{{s_{02}}}}
\end{flalign*}

\begin{flalign*}
&\ {a_{31}} = \frac{{3{c_3}\left( {\varpi _{30}^2 - {c_2} + 1} \right){d_{24}}}}{{2{s_{30}}}},
{a_{32}} =  - \frac{{\left( {\varpi _{30}^2 - {c_2} + 1} \right){t_1} - 2{\varpi _{30}}{t_2}}}{{{s_{30}}}}\\
&\ {a_{33}} =  - \frac{{\left( {\varpi _{30}^2 - {c_2} + 1} \right){t_3} - 3{\varpi _{30}}{t_4}}}{{{s_{30}}}},
{a_{34}} =  - \frac{{\left( {v_{00}^2 - {c_2} + 1} \right){t_5}}}{{{s_{01}}}}\\
&\ {a_{35}} =  - \frac{{\left( {v_{00}^2 - {c_2} + 1} \right){t_6} - 2{v_{00}}{t_7}}}{{{s_{01}}}},
{a_{36}}{\rm{ = }} - \frac{{\left( {\varpi _{21}^2 - {c_2} + 1} \right){t_8}}}{{{s_{21}}}}\\
&\ {a_{37}}{\rm{ = }} - \frac{{\left( {\varpi _{21}^2 - {c_2} + 1} \right){t_9} - 2{\varpi _{21}}{t_{10}}}}{{{s_{21}}}},
{a_{38}} =  - \frac{{\left( {\varpi _{2 - 1}^2 - {c_2} + 1} \right){t_{11}}}}{{{s_{2 - 1}}}}&
\end{flalign*}

\begin{flalign*}
&\ {a_{39}} =  - \frac{{\left( {\varpi _{2 - 1}^2 - {c_2} + 1} \right){t_{12}} - 2{\varpi _{2 - 1}}{t_{13}}}}{{{s_{2 - 1}}}},
{a_{310}} =  - \frac{{\left( {\varpi _{12}^2 - {c_2} + 1} \right){t_{14}}}}{{{s_{12}}}}\\
&\ {a_{311}} = \frac{{\left( {\varpi _{12}^2 - {c_2} + 1} \right){t_{15}} - 2{\varpi _{12}}{t_{16}}}}{{{s_{12}}}},
{a_{312}} =  - \frac{{\left( {\varpi _{1 - 2}^2 - {c_2} + 1} \right){t_{17}}}}{{{s_{1 - 2}}}}\\
&\ {a_{313}} = \frac{{\left( {\varpi _{1 - 2}^2 - {c_2} + 1} \right){t_{18}} - 2{\varpi _{1 - 2}}{t_{19}}}}{{{s_{1 - 2}}}},
{a_{314}} = \frac{{3{c_3}\left( {\varpi _{01}^2 - {c_2} + 1} \right)\left( {2{d_{29}} + {d_{210}}} \right)}}{{2{s_{01}}}}\\
&\ {a_{315}} = \frac{{3{c_3}\left( {\varpi _{01}^2 - {c_2} + 1} \right){d_{210}}}}{{2{s_{03}}}}&
\end{flalign*}

\begin{flalign*}
&\ {b_{21}} = \frac{{3{c_3}}}{2}\frac{{\kappa \varpi _{20}^2 + ({\kappa ^2} + 2){\varpi _{20}} + \kappa (2{c_2} + 1)}}{{{s_{20}}}},
{b_{22}} =  - \frac{{3{c_3}}}{2}\frac{{{\varpi _{20}}}}{{{s_{20}}}}\\
&\ {b_{23}} = \frac{{ - 3{c_3}{\varpi _{11}}}}{{{s_{11}}}},
{b_{24}} = \frac{{ - 3{c_3}{\varpi _{1 - 1}}}}{{{s_{1 - 1}}}},
{b_{25}} =  - \frac{{3{c_3}}}{2}\frac{{{\varpi _{02}}}}{{{s_{02}}}}&
\end{flalign*}

\begin{flalign*}
& 
{b_{31}} =  - \frac{{3{c_3}(\kappa {\omega _0} + 1)(2{d_{22}} + {d_{24}})}}{{2{s_1}}},{b_{32}} = \frac{{(\kappa {\omega _0} + 1){t_{20}} - ({\omega _0} + \kappa ){t_{21}}}}{{{s_1}}}\\
& {b_{33}} = \frac{{(\kappa {\omega _0} + 1){t_{22}} - ({\omega _0} + \kappa ){t_{23}}}}{{{s_1}}},
{b_{34}} =  - \frac{{6{c_3}{\varpi _{30}}{d_{24}}}}{{2{s_{30}}}}\\
& {b_{35}} = \frac{{2{\varpi _{30}}{t_1} - \left( {\varpi _{30}^2 + 2{c_2} + 1} \right){t_2}}}{{{s_{30}}}},
{b_{36}} = \frac{{2{\varpi _{30}}{t_3} - \left( {\varpi _{30}^2 + 2{c_2} + 1} \right){t_4}}}{{{s_{30}}}} \\
& {b_{37}} = \frac{{2{v_{00}}{t_5}}}{{{s_{10}}}},
{b_{38}} = \frac{{2{v_{00}}{t_6} - \left( {v_{00}^2 + 2{c_2} + 1} \right){t_7}}}{{{s_{01}}}}\\
& {b_{39}} = \frac{{2{\varpi _{21}}{t_8}}}{{{s_{21}}}},
{b_{310}} = \frac{{2{\varpi _{21}}{t_9} - \left( {\varpi _{21}^2 + 2{c_2} + 1} \right){t_{10}}}}{{{s_{21}}}}\\
& {b_{311}} = \frac{{2{\varpi _{2 - 1}}{t_{11}}}}{{{s_{2 - 1}}}},
{b_{312}} = \frac{{2{\varpi _{2 - 1}}{t_{12}} - \left( {\varpi _{2 - 1}^2 + 2{c_2} + 1} \right){t_{13}}}}{{{s_{2 - 1}}}}\\
& {b_{313}} = \frac{{(\kappa {\omega _0} + 1){t_{24}}}}{{{s_1}}},
{b_{314}} = \frac{{(\kappa {\omega _0} + 1){t_{25}} - ({\omega _0} + \kappa ){t_{26}}}}{{{s_1}}}\\
& {b_{315}} = \frac{{2{\varpi _{12}}{t_{14}}}}{{{s_{12}}}},
{b_{316}} = \frac{{2{\varpi _{12}}{t_{15}} - \left( {\varpi _{12}^2 + 2{c_2} + 1} \right){t_{16}}}}{{{s_{12}}}}&
\end{flalign*}

\begin{flalign*}
& {b_{317}} = \frac{{2{\varpi _{1 - 2}}{t_{17}}}}{{{s_{1 - 2}}}},{b_{318}} = \frac{{2{\varpi _{1 - 2}}{t_{18}} - \left( {\varpi _{1 - 2}^2 + 2{c_2} + 1} \right){t_{19}}}}{{{s_{1 - 2}}}}\\
& {b_{319}} =  - \frac{{3{c_3}{v_{00}}(2{d_{29}} + {d_{210}})}}{{{s_{01}}}},
{b_{320}} =  - \frac{{3{c_3}{d_{210}}{\varpi _{03}}}}{{{s_{03}}}}&
\end{flalign*}

\begin{flalign*}
& {d_{21}} =  - \frac{{3{c_3}}}{{2{c_2}}} - \frac{{\omega _0^2 - v_0^2}}{{{c_2}}}{a_{21}},{d_{22}} =  - \frac{{\omega _0^2 - v_0^2}}{{{c_2}}}{a_{22}}\\
& {d_{23}} =  - \frac{{3{c_3}}}{{2({c_2} - \varpi _{20}^2)}} - \frac{{\omega _0^2 - v_0^2}}{{{c_2} - \varpi _{20}^2}}{a_{23}},{d_{24}} =  - \frac{{\omega _0^2 - v_0^2}}{{{c_2} - \varpi _{20}^2}}{a_{24}}\\
& {d_{25}} =  - \frac{{3{c_3}}}{{2({c_2} - \varpi _{11}^2)}},{d_{26}} =  - \frac{{\omega _0^2 - v_0^2}}{{{c_2} - \varpi _{11}^2}}{a_{25}}\\
& {d_{27}} =  - \frac{{3{c_3}}}{{2({c_2} - \varpi _{1 - 1}^2)}},{d_{28}} =  - \frac{{\omega _0^2 - v_0^2}}{{{c_2} - \varpi _{1 - 1}^2}}{a_{26}}\\
& {d_{29}} =  - \frac{{3{c_3}(\omega _0^2 - v_0^2)}}{{4(1 + 2{c_2}){c_2}}},{d_{210}} =  - \frac{{\omega _0^2 - v_0^2}}{{{c_2} - \varpi _{02}^2}}{a_{28}}&
\end{flalign*}

\begin{flalign*}
& {d_{31}} =  - \frac{{\omega _0^2 - v_0^2}}{{{c_2} - \varpi _{30}^2}}{a_{31}},{d_{32}} = \frac{{{t_{27}}}}{{{c_2} - \varpi _{30}^2}} - \frac{{\left( {\omega _0^2 - v_0^2} \right){a_{32}}}}{{{c_2} - \varpi _{30}^2}}\\
& {d_{33}} = \frac{{{t_{28}}}}{{{c_2} - \varpi _{30}^2}} - \frac{{\left( {\omega _0^2 - v_0^2} \right){a_{33}}}}{{{c_2} - \varpi _{30}^2}},
{d_{34}} =  - \frac{{\omega _0^2 - v_0^2}}{{{c_2} - \varpi _{21}^2}}{a_{36}}\\
& {d_{35}} = \frac{{{t_{29}} - \left( {\omega _0^2 - v_0^2} \right){a_{37}}}}{{{c_2} - \varpi _{21}^2}},{d_{36}} = \frac{{{t_{30}}}}{{{c_2} - \varpi _{21}^2}},
{d_{37}} =  - \frac{{\omega _0^2 - v_0^2}}{{{c_2} - \varpi _{2 - 1}^2}}{a_{38}}\\
& {d_{38}} = \frac{{{t_{31}}}}{{{c_2} - \varpi _{2 - 1}^2}} - \frac{{\omega _0^2 - v_0^2}}{{{c_2} - \varpi _{2 - 1}^2}}{a_{39}},
{d_{39}} = \frac{{{t_{32}}}}{{{c_2} - \varpi _{2 - 1}^2}}\\
& {d_{310}} =  - \frac{{\omega _0^2 - v_0^2}}{{{c_2} - \varpi _{12}^2}}{a_{310}},{d_{311}} = \frac{{{t_{33}}}}{{{c_2} - \varpi _{12}^2}} - \frac{{\omega _0^2 - v_0^2}}{{{c_2} - \varpi _{12}^2}}{a_{311}}\\
& {d_{312}} =  - \frac{{\omega _0^2 - v_0^2}}{{{c_2} - \varpi _{1 - 2}^2}}{a_{312}},{d_{313}} = \frac{{{t_{34}}}}{{{c_2} - \varpi _{1 - 2}^2}} - \frac{{\omega _0^2 - v_0^2}}{{{c_2} - \varpi _{1 - 2}^2}}{a_{313}}\\
& {d_{314}} =  - \frac{{\omega _0^2 - v_0^2}}{{{c_2} - \varpi _{03}^2}}{a_{315}},{d_{315}} = \frac{3}{8}\frac{{4{c_3}{a_{28}} + {c_4}}}{{{c_2} - \varpi _{03}^2}}&
\end{flalign*}

\begin{flalign*}
& {e_{31}} = \frac{1}{2}\frac{{3{c_3}(\omega _0^2 - {c_2} + 1)(2{d_{22}} + {d_{24}})}}{{2{s_1}}},{e_{32}} =  - \frac{1}{2}\frac{{(\omega _0^2 - {c_2} + 1){t_{20}} - 2{\omega _0}{t_{21}}}}{{{s_1}}}\\
& {e_{33}} =  - \frac{1}{2}\frac{{(\omega _0^2 - {c_2} + 1){t_{22}} - 2{\omega _0}{t_{23}}}}{{{s_1}}},
{e_{34}} = \frac{{\left( {\omega _0^2 - v_0^2} \right){a_{34}}}}{{2{v_{00}}}}&
\end{flalign*}

\begin{flalign*}
& {e_{35}} =  - \frac{{{t_{38}} - \left( {\omega _0^2 - v_0^2} \right){a_{35}}}}{{2{v_{00}}}},{e_{36}} = \frac{{\frac{3}{2}{c_3}\left( {2{a_{21}} + {d_{25}} + {d_{27}}} \right) + \frac{3}{4}\left( {4 - {\kappa ^2}} \right){c_4}}}{{2{v_{00}}}}\\
& {e_{37}} =  - \frac{1}{2}\frac{{(\omega _0^2 - {c_2} + 1){t_{24}}}}{{{s_1}}},{e_{38}} =  - \frac{1}{2}\frac{{(\omega _0^2 - {c_2} + 1){t_{25}} - {\omega _0}{t_{26}}}}{{{s_1}}}\\
& {e_{39}} = \frac{{\omega _0^2 - v_0^2}}{{2{v_{00}}}}{a_{314}},{e_{310}} = \frac{{\frac{3}{2}{c_3}(2{a_{27}} + {a_{28}}) - \frac{9}{8}{c_4}}}{{2{v_{00}}}}&
\end{flalign*}

\begin{flalign*}
& {l_1} =  - 2{\omega _0}{e_{31}},{l_2} =  - 2{\omega _0}{e_{32}} + {t_{35}},{l_3} =  - 2{\omega _0}{e_{34}} - {t_{36}}\\
& {l_4} =  - 2{\omega _0}{e_{33}},{l_5} =  - 2{\omega _0}{e_{35}} - {t_{37}}&
\end{flalign*}
where
\begin{flalign*}
& {\varpi _{km}} = k{\omega _0} + m{v_0}\\
& {s_{km}} = \varpi _{km}^2(\varpi _{km}^2 - 2 + {c_2}) - (2{c_2} + 1)({c_2} - 1)\\
& {s_1} = \omega _0^3 - \kappa {\omega _0}^2 - ({c_2} + 1){\omega _0} - \kappa ({c_2} - 1),&
\end{flalign*}

\begin{flalign*}
& {t_1} = \frac{3}{2}{c_3}\left( {2{a_{24}} + \kappa {b_{22}} - {d_{23}}} \right) - \frac{3}{2}{c_4}\\
& {t_2} =  - \frac{3}{2}{c_3}\left( {\kappa {a_{24}} + {b_{22}}} \right) + \frac{3}{8}{c_4}\kappa \\
& {t_3} = \frac{3}{2}{c_3}\left( {2{a_{23}} + \kappa {b_{21}}} \right) + \frac{1}{2}{c_4}( {2 + 3{\kappa ^2}} )\\
& {t_4} =  - \frac{3}{2}{c_3}\left( {\kappa {a_{23}} + {b_{21}}} \right) - \frac{3}{8}{c_4}\kappa,
{t_5} =  - \frac{3}{2}{c_3}\left( {2{d_{22}} + {d_{26}} + {d_{28}}} \right)\\
& {t_6} = \frac{3}{2}{c_3}\left( {2{a_{25}} + 2{a_{26}} - \kappa {b_{23}} - \kappa {b_{24}} - 2{d_{21}} - {d_{25}} - {d_{27}}} \right) - 6{c_4}\\
& {t_7} = \frac{3}{2}{c_3}\left( {\kappa {a_{25}} - \kappa {a_{26}} - {b_{23}} + {b_{24}}} \right)\\
& {t_8} =  - \frac{3}{2}{c_3}\left( {{d_{24}} + {d_{26}}} \right)&
\end{flalign*}

\begin{flalign*}
& {t_9} = \frac{3}{2}{c_3}\left( {2{a_{25}} + \kappa {b_{23}} - {d_{23}} - {d_{25}}} \right) - 3{c_4}\\
& {t_{10}} =  - \frac{3}{2}{c_3}\left( {\kappa {a_{25}} + {b_{23}}} \right) + \frac{3}{4}\kappa {c_4}\\
& {t_{11}} =  - \frac{3}{2}{c_3}\left( {{d_{24}} + {d_{28}}} \right)\\
& {t_{12}} =  - \frac{3}{2}{c_3}\left( {{d_{23}} + {d_{27}}} \right) - 3{c_4}\\
& {t_{13}} =  - \frac{3}{2}{c_3}\left( {\kappa {a_{26}} + {b_{24}}} \right) - \frac{3}{4}\kappa {c_4}\\
& {t_{14}} =  - \frac{3}{2}{c_3}\left( {{d_{26}} + {d_{30}}} \right)\\
& {t_{15}} = \frac{3}{2}{c_3}\left( {2{a_{28}} + \kappa {b_{25}} - {d_{25}}} \right) - \frac{3}{2}{c_4}\\
& {t_{16}} =  - \frac{3}{2}{c_3}\left( {\kappa {a_{28}} + {b_{25}}} \right) + \frac{3}{8}{c_4}\kappa,
  {t_{17}} =  - \frac{3}{2}{c_3}\left( {{d_{28}} + {d_{30}}} \right)\\
& {t_{18}} = \frac{3}{2}{c_3}\left( {2{a_{28}} - \kappa {b_{25}} - {d_{27}}} \right) - \frac{3}{2}{c_4}\\
& {t_{19}} =  - \frac{3}{2}{c_3}\left( {\kappa {a_{28}} - {b_{25}}} \right) + \frac{3}{8}{c_4}\kappa \\
& {t_{20}} = \frac{3}{2}{c_3}\left( {4{a_{22}} + 2{a_{24}} - \kappa {b_{22}} - 2{d_{21}} - {d_{23}}} \right) - \frac{9}{2}{c_4}\\
& {t_{21}} =  - \frac{3}{2}{c_3}\left( {2\kappa {a_{22}} - \kappa {a_{24}} + {b_{22}}} \right) + \frac{3}{8}{c_4}\kappa\\ 
& {t_{22}} = \frac{3}{2}{c_3}\left( {4{a_{21}} + 2{a_{23}} - \kappa {b_{21}}} \right) + \frac{3}{2}{c_4}( {2 - {\kappa ^2}} )\\
& {t_{23}} =  - \frac{3}{2}{c_3}\left( {2\kappa {a_{21}} - \kappa {a_{23}} + {b_{21}}} \right) - \frac{3}{8}{c_4}\kappa ( {4 - 3{\kappa ^2}} )\\
& {t_{24}} =  - \frac{3}{2}{c_3}\left( {{d_{26}} + {d_{28}} + 2{d_{29}}} \right)\\
& {t_{25}} = \frac{3}{2}{c_3}\left( {4{a_{27}} - {d_{25}} - {d_{27}}} \right) - 3{c_4}\\
& {t_{26}} =  - \frac{3}{4}\kappa \left( {4{c_3}{a_{27}} - {c_4}} \right)\\
& {t_{27}} =  - \frac{3}{2}{c_3}({a_{24}} + {d_{24}}) + \frac{3}{8}{c_4}&
\end{flalign*}

\begin{flalign*}
& {t_{28}} =  - \frac{3}{2}{c_3}\left( {{a_{23}} + {d_{23}}} \right) - \frac{3}{8}{c_4}( {4 + {\kappa ^2}} )\\
& {t_{29}} =  - \frac{3}{2}{c_3}\left( {{a_{24}} + {a_{25}} + {d_{26}}} \right) + \frac{9}{8}{c_4}\\
& {t_{30}} =  - \frac{3}{2}{c_3}\left( {{a_{23}} + {d_{25}}} \right) - \frac{3}{8}{c_4}( {4 + {\kappa ^2}} )\\
& {t_{31}} =  - \frac{3}{2}{c_3}\left( {{a_{24}} + {a_{26}} + {d_{28}}} \right) + \frac{9}{8}{c_4}\\
& {t_{32}} =  - \frac{3}{2}{c_3}\left( {{a_{23}} + {d_{27}}} \right) - \frac{3}{8}{c_4}( {4 + {\kappa ^2}} )\\
& {t_{33}} =  - \frac{3}{2}{c_3}\left( {{a_{25}} + {a_{28}} + {d_{210}}} \right) + \frac{9}{8}{c_4}\\
& {t_{34}} =  - \frac{3}{2}{c_3}\left( {{a_{26}} + {a_{28}} + {d_{210}}} \right) + \frac{9}{8}{c_4}\\
& {t_{35}} =  - \frac{3}{2}{c_3}\left( {2{a_{22}} + {a_{24}} + 2{d_{22}} + {d_{24}}} \right) + \frac{9}{8}{c_4}\\
& {t_{36}} =  - \frac{3}{2}{c_3}\left( {2{a_{21}} + {a_{23}} + 2{d_{21}} + {d_{23}}} \right) - \frac{3}{8}{c_4}( {12 - {\kappa ^2}} )\\
& {t_{37}} =  - \frac{3}{2}{c_3}\left( {{a_{25}} + {a_{26}} + 2{a_{27}} + 2{d_{29}}} \right) + \frac{9}{4}{c_4}\\
& {t_{38}} =  - \frac{3}{2}{c_3}\left( {2{a_{22}} + {a_{25}} + {a_{26}} + {d_{26}} + {d_{28}}} \right) + \frac{9}{4}{c_4}.&
\end{flalign*}

\newpage

\section{Coefficient of expansion for center manifolds of {$L_1$} in Sun-Earth system (up to order 3)}
\label{B:numerical Coefficients}
\begin{table*}[htp]
    \centering
    \tiny
    \begin{tabular}{cccclll}
        \hline
         $\boldsymbol{i}$ & $\boldsymbol{j}$ & & & $\omega_{ij}$ & $v_{ij}$ & $d_{ij}$ \\
         \hline
         0&	0& & &	2.0864535642231&	2.01521066299663&	-0.292214459403954 \\
         2&	0& & &	\makecell*[l]{ -1.72061652811836 \\ +0.190350147100190$\eta ^2$ \\ -0.00435734903357697$\eta ^4$ } & 
         \makecell*[l]{ 0.222743075098847 \\ -0.787717968928588$\eta ^2$ \\ 0.00713914812459876$\eta ^4$ } & 
         \makecell*[l]{ 13.7987585114454 \\ -1.63237220178359$\eta ^2$ \\ +0.0181828128433413$\eta ^4$ } \\	         
         0&	2& & &	\makecell*[l]{ 0.0258184143757671 \\ -0.00866849848354153$\eta ^2$ } & 
         \makecell*[l]{ -0.163191575817707 \\ 0.00354869445928051$\eta ^2$ } & 
         \makecell*[l]{ -1.61744593710231 \\ +0.0361728391148951$\eta ^2$ }	\\
         \hline
         $\boldsymbol{i}$ & $\boldsymbol{j}$ & $\boldsymbol{k}$ & $\boldsymbol{m}$ & $x_{ijkm}$ & $y_{ijkm}$ & $z_{ijkm}$ \\
         \hline
         1&	0&	1&	0&	1&	-3.22926825193629&	$\eta$ \\
         0&	1&	0&	1&	1& &		\\
         2&	0&	0&	0&	\makecell*[l]{ 2.09269572450663 \\ +0.248297657691632$\eta ^2$} & & 	
                         \makecell*[l]{-1.26605225820339$\eta$  \\  -0.0178662505345158$\eta ^3$} \\
        2&	0&	2&	 0&  \makecell*[l]{-0.905964830191359 \\ 0.104464108531470$\eta ^2$ } & 
                         \makecell*[l]{-0.492445878382662 \\ -0.0607464599707783$\eta ^2$}	&
                         \makecell*[l]{0.319446857147281$\eta$ \\ 0.00228622980549827$\eta ^3$} \\

        0&	2&	0&	 0&  0.248297657691632& &		-0.0178662505345158$\eta$  \\
        
        0&	2&	0&	 2&  0.110825182204290&	-0.0677637342617734&	0.00265814089052512$\eta$  \\
        
        1&	1&	1&	-1&  0.495958173029419$\eta$&	0.0231240293704513$\eta$&	
                         \makecell*[l]{-1.11686826756838 \\ -0.0357313126864700$\eta ^2$} \\
                         
        1&	1&	1&	 1&  0.215140142107999$\eta$&	-0.128236554280252$\eta$&	
                         \makecell*[l]{0.354945285830462 \\ +0.00492589138712682$\eta ^2$} \\
        
        3&	0&	1&	 0&  & \makecell*[l]{2.84508162474333 \\ -0.121704813945821$\eta ^2$} &  \\
        
        3&	0&	3&	 0&  \makecell*[l]{-0.793820244082386 \\ +0.0798601114091475$\eta ^2$ \\ +0.000235578066745959$\eta ^4$} &
            \makecell*[l]{-0.885700891209062 \\ 0.0239990788365673$\eta ^2$ \\ -8.16516290582984e-05$\eta ^4$} &
            \makecell*[l]{0.384640956092706$\eta$ \\ -0.0179260040244910$\eta ^3$ \\ 1.96019971748180e-06$\eta ^5$}  \\
        
        1&	2&	1&	-2&  \makecell*[l]{-1.49999489157672 \\ 0.0183747626073067$\eta ^2$} &
                         \makecell*[l]{-4.84196804175048 \\ 0.0995072378868759$\eta ^2$} & 
                         \makecell*[l]{3.85337485129577$\eta$ \\ -0.0190360469734282$\eta ^3$} \\
                         
        1&	2&	1&	 0&  & \makecell*[l]{0.287553231581211 \\ -0.0769801043821544$\eta ^2$} &  \\
        
        1&	2&	1&	 2&  \makecell*[l]{0.0838777765981095 \\ +0.000814467474446631$\eta ^2$} &
                         \makecell*[l]{0.0208288184463949 \\ -0.000290029597041327$\eta ^2$} & 
                         \makecell*[l]{-0.0568481703360723$\eta$ \\ 7.13531915699632e-06$\eta ^3$}\\
                         
        2&	1&	2&	-1&  \makecell*[l]{0.386666472970278$\eta$ \\ -0.0721768185057328$\eta ^3$} &
                         \makecell*[l]{-0.927668808967796$\eta$ \\ + 0.195331264846977$\eta ^3$} & 
                         \makecell*[l]{12.1656581373461 \\ -1.24172810236909$\eta ^2$ \\ -0.0354722817195465$\eta ^4$} \\
                         
        2&	1&	2&	 1&  \makecell*[l]{0.163761849329643$\eta$ \\ 0.000758601954199828$\eta ^3$} &
                         \makecell*[l]{0.0449861418103874$\eta$ \\ -0.000266482431864747$\eta ^3$} &  
                         \makecell[l]{0.406079303697784 \\ -0.0552521061769830$\eta ^2$}\\
                         
        2&	1&	0&	 1&  \makecell*[l]{-5.77468672378054$\eta$ \\ 0.0984680050039235$\eta ^3$} &
                         \makecell*[l]{18.4807696836236$\eta$ \\ -0.396867547295820$\eta ^3$} &  \\
                         
        0&	3&	0&	 1& 0.0489460167621145$\eta$ &	-0.197273069780448$\eta$ &	 \\
        
        0&	3&	0&	 3& 0.000291516143361502$\eta$ &	-0.000105253712354915$\eta$ &	
        \makecell*[l]{-0.0195272217510433\\2.62200442231958e-06$\eta ^2$} \\

         \hline
    \end{tabular}
\end{table*}

\newpage

 \bibliographystyle{elsarticle-num} 
 \bibliography{cas-refs-lin}





\end{document}